\documentclass[12pt,draftcls, onecolumn]{IEEEtran}

\usepackage{amsmath}
\usepackage{amssymb}
\usepackage[dvips]{graphicx}
\usepackage{setspace}
\usepackage{epsfig}

\newcommand{\K}{{\sf{K}}}

\newcommand{\y}{\mathbf{y}}

\newcommand{\R}{{\mathbf{R}}}

\newcommand{\snr}{{\footnotesize{SNR}}}
\newcommand{\tsnr}{{\text{\footnotesize{SNR}}}}

\newcommand{\ud}{{\mathrm{d}}}
\newcommand{\D}{{\mathcal{D}}}
\newcommand{\htau}{{\hat{\tau}}}
\newcommand{\ttau}{{\tilde{\tau}}}

\newtheorem{prop:oopskc}{Proposition}
\newtheorem{prop:oofskc}[prop:oopskc]{Proposition}
\newtheorem{prop:oopsknc}[prop:oopskc]{Proposition}
\newtheorem{prop:oofsknc}[prop:oopskc]{Proposition}

\newtheorem{rem:decisionregionshape}{Remark}
\newtheorem{rem:avgerrorprob}[rem:decisionregionshape]{Remark}
\newtheorem{rem:lowasymptotics}[rem:decisionregionshape]{Remark}
\newtheorem{rem:highasymptotics}[rem:decisionregionshape]{Remark}
\newtheorem{rem:avgerrorproboofsk}[rem:decisionregionshape]{Remark}
\newtheorem{rem:oofsk}[rem:decisionregionshape]{Remark}
\newtheorem{rem:drsoopskn}[rem:decisionregionshape]{Remark}
\newtheorem{rem:lowasymptoticsoopskn}[rem:decisionregionshape]{Remark}
\newtheorem{rem:highasymptoticsoopskn}[rem:decisionregionshape]{Remark}
\newtheorem{rem:asymptoticsoofskn}[rem:decisionregionshape]{Remark}
\newtheorem{rem:bounds}[rem:decisionregionshape]{Remark}

\begin{document}

\title{Error Rate Analysis for Peaky Signaling over Fading Channels \footnote{This work was supported in part by the NSF CAREER Grant
CCF-0546384. The material in this paper was presented in part at the
40th Annual Asilomar Conference on Signals, Systems, and Computers,
Nov. 2006, and the 40th Annual Conference on Information Sciences
and Systems, Princeton University, Princeton, NJ, March, 2006.}}

%\author{\authorblockN{Michael Shell} \and
%\and \authorblockN{James Kirk\\ and Montgomery Scott}
%\authorblockA{Starfleet Academy\\
%San Francisco, California 96678-2391\\ Telephone: (800)
%555--1212\\ Fax: (888) 555--1212}}

% avoiding spaces at the end of the author lines is not a problem with
% conference papers because we don't use \thanks or \IEEEmembership

{\small{
\author{\authorblockN{Mustafa Cenk Gursoy}
\\
\authorblockA{Department of Electrical Engineering\\
University of Nebraska-Lincoln, Lincoln, NE 68588\\ Email:
gursoy@engr.unl.edu}}}}

%\specialpapernotice{(Invited Paper)}

\maketitle \vspace{-1.5cm}
\begin{abstract} In
this paper, the performance of signaling strategies with high
peak-to-average power ratio is analyzed over both coherent and
noncoherent fading channels.  Two modulation schemes, namely on-off
phase-shift keying (OOPSK) and on-off frequency-shift keying
(OOFSK), are considered. Initially, uncoded systems are analyzed.
For OOPSK and OOFSK, the optimal detector structures are identified
and analytical expressions for the error probabilities are obtained
for arbitrary constellation sizes. Numerical techniques are employed
to compute the error probabilities. It is concluded that increasing
the peakedness of the signals results in reduced error rates for a
given power level and hence equivalently improves the energy
efficiency for fixed error probabilities. The coded performance is
also studied by analyzing the random coding error exponents achieved
by OOPSK and OOFSK signaling.


\end{abstract}

\begin{spacing}{1.7}

\section{Introduction}

In wireless communications, when the receiver and transmitter have
only imperfect knowledge of the channel conditions, efficient
transmission strategies have a markedly different structure than
those employed over perfectly known channels. For instance,
Abou-Faycal \emph{et al.} \cite{Abou} studied the noncoherent
Rayleigh fading channel where the receiver and transmitter has no
channel side information, and showed that the capacity-achieving
input amplitude distribution is discrete with a finite number of
mass points. It has also been shown that there always exists a mass
point at the origin. These results indicate that unlike perfectly
known channels where a Gaussian input is optimal and a continuum of
amplitude levels are available for tranmission, only finitely many
amplitude levels with one level at the origin should be used for
transmission over noncoherent Rayleigh channels. The discreteness of
the optimal amplitude distribution has also been shown over other
noncoherent channels (see e.g., \cite{Chan}, \cite{Meyn}, \cite{Katz
Shamai}, \cite{gursoy1}, and references therein).

Another key result for noncoherent channels is the requirement of
transmission with high peak-to-average power ratio in the low
signal-to-noise ratio (SNR) regime. In noncoherent Rayleigh channels
\cite{Abou}, the optimal amplitude has two mass points for low SNR
values, and the nonzero mass migrates away from the origin,
increasing the peak power, as SNR decreases. Indeed, this behavior
has been shown in a more general setting in \cite{Verdu} where
\emph{flash signaling} is proven to be the necessary form of
transmission to achieve the capacity of imperfectly-known fading
channels in the low-SNR regime.

The impact upon the channel capacity of using signals with limited
peakedness is investigated in \cite{gursoy2}, \cite{Medard
Gallager}, \cite{Subramanian Hajek}, and \cite{Telatar Tse}. In
\cite{gursoy2}, two types of signaling schemes are defined: on-off
binary phase-shift keying (OOBPSK) and on-off quaternary phase-shift
keying (OOQPSK). These modulations are obtained by overlaying on-off
keying on phase-shift keying. The peakedness of these signals are
controlled by changing the probability of no transmission. OOQPSK is
shown to be an optimally efficient modulation scheme for
transmission over noncoherent Rician fading channels in the low-SNR
regime. In \cite{Gursoy}, the capacity and power efficiency of
another type of peaky signaling scheme, on-off frequency-shift
keying (OOFSK), have been studied in Rician fading channels with
perfect or imperfect channel information at the receiver. Note that OOFSK
introduces peakedness in both time and frequency.%and it is
%shown that the minimum bit
%   energy decreases if signals with higher peak-to-average power
%   ratios are allowed.

The aforementioned studies have focused on the achievable rates and
channel capacity. Recently, there have also been interest in the
error performance in noncoherent fading channels. Reference
\cite{gursoy_icc06} considered the error exponents and cutoff rate
of noncoherent Rician fading channels and showed that the optimal
input again has a discrete structure. Huang \emph{et al}
\cite{HuangMeynMedard} proved, under more general assumptions on the
channel, the discreteness of the input that achieves the optimal
random coding error error exponent. Wu and Srikant \cite{Wu}
considered multiple-input multiple-output (MIMO) fading channels and
characterized the reliability function in the low-$\tsnr$ regime.
They analyzed the relation between the communication rate, error
exponents, number of antennas, and signal peakiness. Ray \emph{et
al.} \cite{Ray} analyzed the capacity and the random coding error
exponents of noncoherent Rayleigh MIMO fading channels in the
wideband regime in which the $\tsnr$ per degree of freedom is low
and coherence length is large. Lun \emph{et al.} \cite{DesmondLun}
also investigated the error probabilities of peaky signaling. They
considered peaky FSK modulation and characterized the reliability
function in the infinite bandwidth regime. Luo \emph{et al.}
\cite{ChengLuo} introduced multitone FSK schemes and obtained upper
and lower bounds on the probability of error of these signaling
schemes over wideband noncoherent Rayleigh fading channels.

The above-mentioned studies mainly analyze the error probability of
coded systems in noncoherent fading channels by considering the
error exponents. Generally, the analysis is performed in the
low-$\tsnr$ regime. \cite{gursoy_icc06} and \cite{HuangMeynMedard}
consider general $\tsnr$ levels but focus on the structure of the
input distribution that achieves the maximum error exponents. In
this paper, we present a different approach. Our contributions are
the following. We consider two particular peaky modulation schemes:
on-off phase-shift keying (OOPSK) and on-off frequency-shift keying
(OOFSK). We determine the optimal detector structures and analyze
the error probabilities of uncoded OOPSK and OOFSK.  The analysis is
conducted for both coherent and noncoherent fading channels, and
error performances are investigated at low-to-high $\tsnr$ levels.
We find that the error performance of signaling with high
peak-to-average power ratio is superior to that of conventional PSK
and FSK modulations over the entire $\tsnr$ range if the duty cycles
of modulations are small enough. Equivalently, peaky signaling is
shown to be more energy efficient for fixed error probabilities. We
have also considered the coded performance by obtaining the error
exponents of OOPSK and OOFSK modulations.
%at both low and medium
%$\tsnr$ levels.
%Overall, in the paper, the performance gains of
%employing these information-theory inspired signaling schemes are
%quantified.
As a result of the analysis conducted in this paper,
information-theoretic inspired signaling schemes, OOPSK and OOFSK,
emerge as energy efficient modulation formats especially well-suited
for low data rate applications such as in sensor networks.

%Potential gains in error performance are anticipated in both
%coherent and noncoherent fading channels as increasing the PAR
%without changing the average power consumption increases the minimum
%distance between the constellation points. This is achieved by
%increasing the probability of no signal transmission, which
%potentially decreases the data rate. One possible application in
%which such transmission techniques are desirable is wireless sensor
%networks where low-data rate
%transmissions with low energy consumption are required.
%For
%instance, the specifications of a wireless microsensor system given
%in \cite{Wang} indicate that the microsensors need to operate for
%5-10 years with one AA size battery.

The rest of the paper is organized as follows. We describe the
modulation techniques and the channel model in Section
\ref{sec:model}. We study the error probabilities in coherent fading
channels in Section \ref{sec:coherent} while the performance in
noncoherent fading channels is analyzed in Section
\ref{sec:noncoherent}. We study the random coding error exponents of
peaky signaling in Section \ref{sec:exponents}. Finally, Section
\ref{sec:conclusion} includes our conclusions.

\section{System Model} \label{sec:model}
In this section, we present the system model. We consider two types
of signaling at the transmitter side, OOPSK and OOFSK. Basically,
these two modulation schemes are obtained by overlaying on-off
keying on phase-shift keying and frequency-shift keying,
respectively. In both signaling schemes, transmitter sends over the
symbol interval of $[0, T]$ either no signal with probability
$1-\nu$ or one of $M$ signals, each with probability $\nu/M$. The
transmitted signal can be mathematically expressed as
\begin{align}\label{eq:signal}
s_i(t)= \left\{
\begin{array}{lll}
\sqrt{\frac{P}{\nu}}\, e^{j(\omega_it+\theta_i)} & 1\leq i\leq M &
\text{with prob. } \nu/M
\\
0  & i = 0 & \text{with prob. } (1-\nu)
\end{array}\right. \quad 0 \le t \le
T
\end{align}
where $P$ is the average power, $\omega_i$ and $\theta_i$ are the
frequency in radians and phase of $s_i(t)$, respectively. $s_0(t) =
0$ denotes no transmission. In OOPSK modulation, the frequency is
fixed, i.e., $\omega_i = \omega$ $\forall i$, and phases are
$\theta_i = \frac{2\pi (i-1)}{M}$ for $1 \le i \le M$. In OOFSK,
information is carried by the frequencies and each nonzero signal
has a distinct frequency. To ensure orthogonality, adjacent
frequency slots satisfy $|\omega_{i+1}-\omega_i| = \frac{2\pi}{T}$.
In OOFSK, phases can be arbitrary. In both modulations, $\nu$ can be
regarded as the duty cycle of the transmission. Note also that both
modulation formats have an average power of $P$ and a peak power of
$\frac{P}{\nu}$, and hence a peak-to-average power ratio (PAR) of
$\frac{1}{\nu}$. Limitations on the PAR of the signaling scheme,
which may be dictated by regulations or system component
specifications, are reflected in the choice of the value of $\nu$.

We assume that the transmitted signal undergoes stationary and
ergodic fading and that the delay spread of the fading is much less
than the symbol duration. Under this narrowband assumption, the
fading has a multiplicative effect on the transmitted signal. If we
further assume that the symbol duration $T$ is less than the
coherence time of the fading, then the fading stays constant over
the symbol duration. Hence, if the transmitted signal is $s_i(t)$,
the received signal is
\begin{equation}\label{model}
r(t) = h_k\, s_i(t - (k-1)T) + n(t), \quad (k-1)T \le t \le kT,
\quad \text{ for } i = 0,1,2, \ldots, M \text{ and }k = 1,2, \ldots,
\nonumber
\end{equation}
where $\{h_k\}$ denotes the sequence of fading coefficients and is a
proper, complex, stationary, ergodic fading process with finite
variance, and $n(t)$ is a zero-mean, circularly symmetric, white
complex Gaussian noise process with single-sided spectral density
$N_0$. The transmitted signal, fading coefficients, and additive
noise are assumed to be mutually independent of each other.

If OOPSK modulation is used at the transmitter, the receiver
demodulates the received signal using a correlator: \vspace{-.28cm}
\begin{align}\label{eq:demodoopsk}
y_k&= \frac{1}{\sqrt{N_0T}}\int_{(k-1)T}^{kT}{r(t)e^{-j\omega (t - (k-1)T)}}dt \\
&=\frac{1}{\sqrt{N_0T}}\int_0^T\left[h_k s_i(t)e^{-j\omega t}+n(t +
(k-1)T)e^{-j\omega t}\right]dt =\left\{\begin{array}{ll}\alpha h_k
e^{j\theta_i}+n_k & 1 \le i \le M\\
n_k & i = 0 \end{array}\right. k = 1,2,3 \ldots \nonumber
\end{align}
where $\alpha = \sqrt{\frac{PT}{\nu N_0}}$, $\theta_i = \frac{2\pi
(i-1)}{M}$ for $1 \le i \le M$, and $\{n_k\}$ is a sequence of
independent and identically distributed (i.i.d.) circularly
symmetric complex Gaussian random variables with zero mean and
variance $E\{|n_k|^2\} = 1$.

If OOFSK signals are transmitted, a bank of $M$ correlators is
employed at the receiver and the output of the $m^{th}$ correlator
at time $t=kT$ is given by:
\begin{align}\label{eq:output}
y_{k,m} &= \frac{1}{\sqrt{N_0T}}\int_{(k-1)T}^{kT}{r(t)e^{-j\omega_m
(t -
(k-1)T)}}dt\\
&=\frac{1}{\sqrt{N_0T}}\int_0^T\left[h_k s_i(t)e^{-j\omega_m t}+n(t
+ (k-1)T)e^{-j\omega_m t}\right]dt
\\
&=\left\{\begin{array}{ll}\alpha h_k
e^{j\theta_i}+n_{k,m} &m=i\\
n_{k,m}  &m \neq i \end{array},\right. \quad m = 1,2,\ldots, M,
\quad i= 0,1,\ldots,M \quad k=1,2,\ldots \label{eq:outputoofsk}
\end{align}
where again $\alpha = \sqrt{\frac{PT}{\nu N_0}}$, and $\{n_{k,m}\}$
is an i.i.d sequence in both $k$ and $m$ of zero-mean unit-variance
circularly symmetric complex Gaussian random variables. The output
of $M$ demodulators is denoted by the $M$-dimensional vector $\y_k =
[y_{k,1}, y_{k,2}, \ldots, y_{k,M}]$.

\section{Error Probability in Coherent Fading Channels}
\label{sec:coherent}

In this section, we assume that the receiver perfectly knows the
instantaneous realizations of the fading coefficients $\{h_k\}$
whereas the transmitter has no such knowledge.

%Our focus will be on uncoded systems so that error performance
%improvements obtained by using peaky signals can be identified in
%the absence of coding gain achievements.

\subsection{OOPSK} \label{subsec:coopsk}

We first consider OOPSK signaling. For the detection of these
signals, maximum a posteriori probability (MAP) decision rule, which
minimizes the probability of error, is employed after demodulation
at the receiver. It is assumed that symbol-by-symbol detection is
performed, and henceforth we drop the time index $k$ without loss of
generality. In MAP detection, signal $s_i$ is chosen as the detected
signal if \vspace{-.2cm}
\begin{gather} \label{eq:map}
p_i \, f_{y | s_i, h}(y | s_i, h) > p_j \, f_{y | s_j, h}(y | s_j,
h) \quad \forall j \neq i
\end{gather}
where $p_i$ and $p_j$ denote the prior transmission probabilities of
the signals $s_i$ and $s_j$, respectively, and
\begin{gather} \label{eq:cpdf}
f_{y | s_i, h}(y | s_i, h) = \left\{
\begin{array}{ll}
\frac{1}{\pi} e^{-|y - \alpha h e^{j\theta_i}|^2} & 1 \le i \le M
\\
\frac{1}{\pi} e^{-|y|^2} & i = 0
\end{array}\right.
\end{gather}
is the conditional probability density function of the output given
$s_i$ and $h$. Note from (\ref{eq:signal}) that $p_i = \nu/M$ for $i
\neq 0$ and $p_0 = 1-\nu$. The following result provides the
decision regions and the error probability for OOPSK modulation.
\begin{prop:oopskc}
The optimal decision regions for OOPSK signals when transmitted over
coherent fading channels are
\begin{gather}\label{eq:droopsk}
\D_i = \left\{ y = |y|e^{j\theta_y}: \frac{2\pi (i-\frac{3}{2})}{M}
\le \theta_y \le \frac{2\pi (i-\frac{1}{2})}{M} \text{ and } |y|
\cos(\theta_y - \theta_i) > \tau\right\} \quad 1\le i \le M,
\\
\D_0 = \left\{y = |y|e^{j\theta_y}: |y| \cos(\theta_y - \theta_i) <
\tau \quad \forall i \neq 0\right\},
\end{gather}
where $\tau = \left\{
\begin{array}{ll}
\zeta & \zeta \ge 0
\\
0 & \zeta < 0
\end{array}\right.$
and $ \zeta = \frac{\alpha|h|}{2} + \frac{1}{2 \alpha |h|} \ln
\left( \frac{M(1-\nu)}{\nu}\right) $. Moreover, the error
probability as a function of the instantaneous realization of the
fading  magnitude, $|h|$, is
\begin{gather}
P_{e | \, |h|} = 1 - \left((1-\nu)P_{c|s_0,|h|} + \nu
P_{c|s_1,|h|}\right), \label{eq:errorprob}
\end{gather}
\vspace{-1.3cm} \\
where
\begin{align}
P_{c| s_0, |h|}= M \int_0^\tau \!\!\left(1- 2Q\left(\sqrt{2} x \tan
\frac{\pi}{M}\right)\right) \frac{e^{-x^2}}{\sqrt{\pi}}  \, dx
\text{ and } P_{c|s_1,|h|} = \int_{\tau}^{\infty} \!\!\left(1 -
2Q\left(\sqrt{2}\,x \tan\frac{\pi}{M}\right) \right) \frac{e^{-(x
-\alpha |h|)^2}}{\sqrt{\pi}}  \, dx \label{eq:correctprob}
\end{align}
are the correct detection probabilities when $s_0(t)$ and $s_1(t)$,
respectively, are the transmitted signals.
\end{prop:oopskc}
\emph{Proof}: Using the property that phase-shift keying signals
have the same energy, the detection rule in (\ref{eq:map}) can
easily be simplified to the following: The signal $s_i$ for $i \neq
0$ is the detected signal if
\begin{gather}
\Re(y e^{-j\theta_i})> \Re(y e^{-j\theta_j}) \quad \forall j \neq i,
0 \quad \text{and} \quad \Re(y e^{-\theta_i}) > \tau
\label{eq:smap1}
\end{gather}
where $\Re(z)$ is used to denote the real part of the complex scalar
$z$, and $\tau$ is defined in the proposition above. No transmission
is detected if
\begin{gather} \label{eq:smap0}
\Re(y e^{-j\theta_i}) < \tau  \quad \forall i \neq 0.
\end{gather}
Note that the phase $\theta_h$ of the fading coefficients has no
effect on the error probability as long as the receiver, knowing
perfectly the phase rotations introduced by the channel, removes any
phase offset by using $\tilde{y} = y e^{-\theta_h}$ at the detector.
Hence, the detection rules (\ref{eq:smap1}) and (\ref{eq:smap0}) are
obtained by assuming without loss of generality that $\theta_h = 0$
and hence $h = |h|$. These detection rules lead to the decision
regions given in the proposition. After the identification of the
decision regions, performance analysis is easily conducted. The
correct detection probabilities in (\ref{eq:correctprob}) are
obtained from
\begin{gather}
\!\!\!\!\!P_{c| s_0, h} = P(y \in D_0 | s_0, h) =
 M P\left(y_r < \tau, |y_i| < y_r \tan
\frac{\pi}{M}| s_0, h\right) = M \int_0^\tau \int_{-y_r \tan
\frac{\pi}{M}}^{y_r \tan \frac{\pi}{M}} f_{y|s_0,h}(y|s_0,h) \, dy,
\\
\!\!\!\!\!\!\!P_{c|s_1,h} = P(y \in \D_1|s_1,h) = P\left(y_r > \tau,
|y_i| < y_r \tan \frac{\pi}{M}|s_1,h\right) = \int_\tau^\infty
\int_{-y_r \tan \frac{\pi}{M}}^{y_r \tan \frac{\pi}{M}}
f_{y|s_1,h}(y|s_1,h) \, dy, \text{ and } \label{eq:Pc1}
\end{gather}
respectively, by expressing
the inner integrals using the Gaussian $Q$-function. In the above
formulation, we have defined $y = y_r + j y_i$. Note that due to the
circular symmetry of the constellation and the decision regions, the
correct detection probabilities of the nonzero signals other than
$s_1(t)$ are the same as in (\ref{eq:Pc1}). Hence, the error
probability can be expressed as in (\ref{eq:errorprob}). \hfill
$\square$

\begin{rem:decisionregionshape}
 For $M \ge 3$, we can easily observe that the region $D_0$
is an $M$-sided regular polygon\footnote{A regular polygon is an
$n$-sided convex polygon in which the sides are all the same length
and are symmetrically placed about a common center.} centered at the
origin.  Therefore, the nonzero signal $s_i(t)$ is detected if the
received signal phase $\theta_y$ is closest to $\theta_i$ and $y$ is
outside the regular polygon for $M \ge 3$. For $M = 2$, $\D_0$ is an
infinite rectangle.
\end{rem:decisionregionshape}

\begin{rem:avgerrorprob}
The probability of error averaged over the realizations of the
fading magnitude is obtained from $P_e = \int_0^\infty P_{e| \, |h|}
\, d F_{|h|}(|h|)$ where $F_{|h|}$ is the distribution function of
the fading magnitude.
\end{rem:avgerrorprob}

\begin{rem:lowasymptotics}\label{rem:lowasymptotics} As $\tsnr \to 0$, we have $\alpha
\to 0$, and hence
\begin{gather}
\zeta \to \left\{
\begin{array}{ll}
\infty & \nu < \frac{M}{M+1}
\\
-\infty & \nu > \frac{M}{M+1}
\\
0 & \nu = \frac{M}{M+1}
\end{array}\right. \text{ and }\,\,\,
\tau \to \left\{
\begin{array}{ll}
\infty & \nu < \frac{M}{M+1}
\\
0 & \nu \ge \frac{M}{M+1}
\end{array}\right..
\end{gather}
If $\nu < \frac{M}{M+1}$, the threshold $\tau$ increases without
bound as $\tsnr$ vanishes. Therefore, $\D_0$ asymptotically becomes
the complex plane and we have $P_{c|s_0,|h|} \to 1$, $P_{c|s_1,|h|}
\to 0$, and $P_{e||h|} \to \nu$. On the other hand, if $\nu \ge
\frac{M}{M+1}$, the decision region $\D_0$ vanishes and
$P_{c|s_0,|h|} \to 0$. Indeed, for sufficiently small values of
$\tsnr$ for which $\zeta < 0$, we have $\tau = 0$. In such a case,
$s_0(t)$ is never detected and the decision regions specialize to
the decision regions of regular PSK modulation.
%because $|y|
%\cos(\theta_y - \theta_i) \ge \tau = 0$ is always satisfied.
\end{rem:lowasymptotics}

\begin{rem:highasymptotics} \label{rem:highasymptotics}
As $\tsnr \to \infty$, it can be easily seen that $\tau \to \infty$.
Hence, clearly $P_{c|s_0,|h|} \to 1$. By applying a change of
variables with $\hat{x} = x - \alpha|h|$, we can express the correct
detection probability of $s_1(t)$ in (\ref{eq:correctprob}) as
\begin{gather}
P_{c|s_1,|h|} = \int_{\tau-\alpha |h|}^{\infty} \!\!\left(1 -
2Q\left(\sqrt{2}\,(\hat{x} + \alpha|h|) \tan\frac{\pi}{M}\right)
\right) \frac{e^{-\hat{x}^2}}{\sqrt{\pi}}  \, d\hat{x} \quad
\substack{\longrightarrow \\ \tsnr \to \infty} \quad
\int_{-\infty}^{\infty} \!\! \frac{e^{-\hat{x}^2}}{\sqrt{\pi}}  \,
d\hat{x} = 1. \label{eq:highasymptotics}
\end{gather}
The limiting value on the right hand side of
(\ref{eq:highasymptotics}) is obtained by noting that as $\alpha \to
\infty$, $\tau - \alpha |h| = -\frac{\alpha|h|}{2} + \frac{1}{2
\alpha |h|} \ln \left( \frac{M(1-\nu)}{\nu}\right) \to -\infty$ and
$Q\left(\sqrt{2}\,(\hat{x} + \alpha|h|) \tan\frac{\pi}{M}\right) \to
0$. Therefore, not surprisingly, $P_{e|h|} \to 0$ as $\tsnr$
increases.
\end{rem:highasymptotics}

\begin{figure}
\begin{center}
\includegraphics[width = 0.65\textwidth]{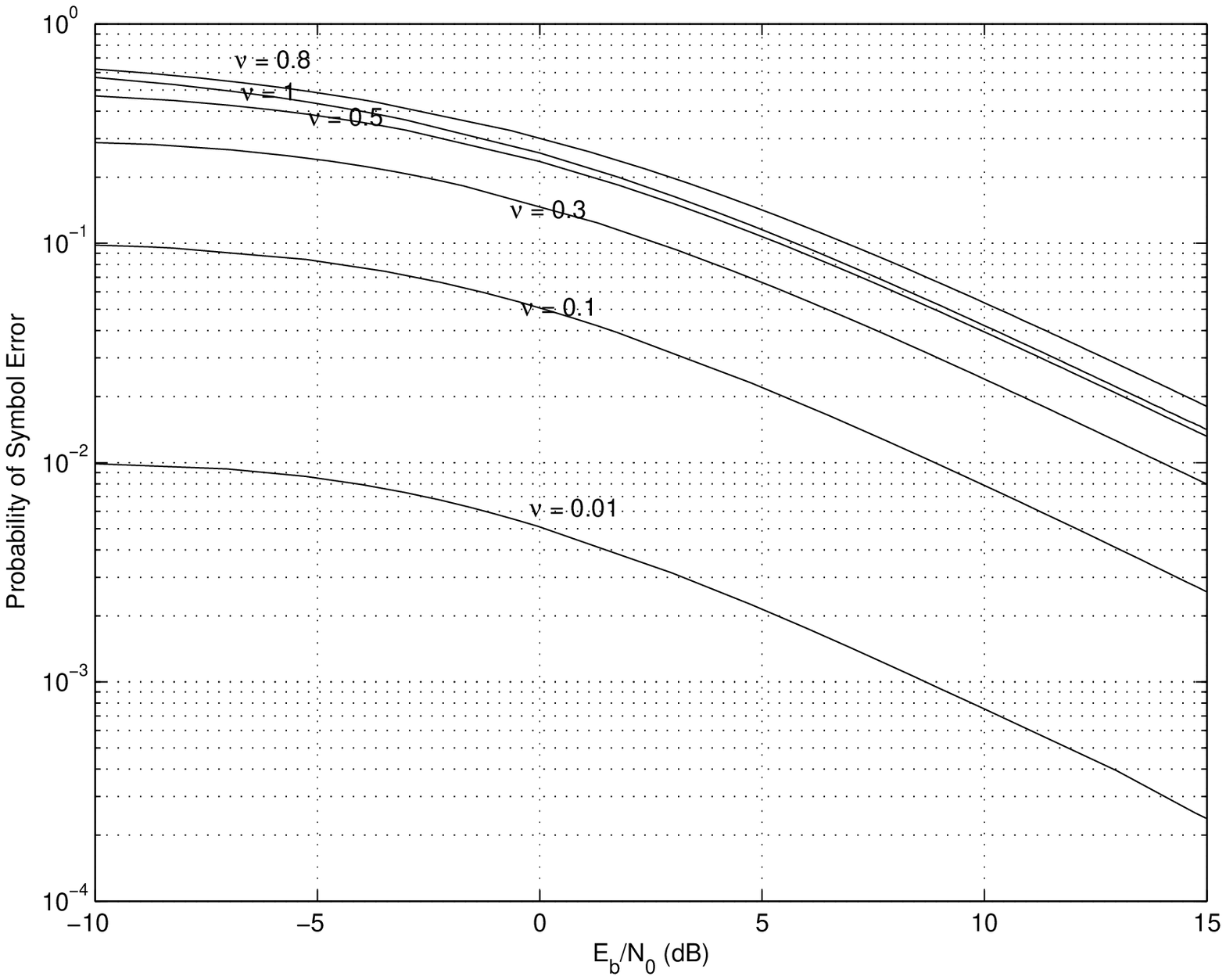}
\caption{Probability of symbol error vs. $E_b/N_0$ for 4-OOPSK
signaling with different duty factor values, $\nu$, in coherent
Rayleigh fading channels with $E\{|h|^2\} = 1$.} \label{fig:oopskM4}
\end{center}
\end{figure}

\begin{figure}
\begin{center}
\includegraphics[width = 0.65\textwidth]{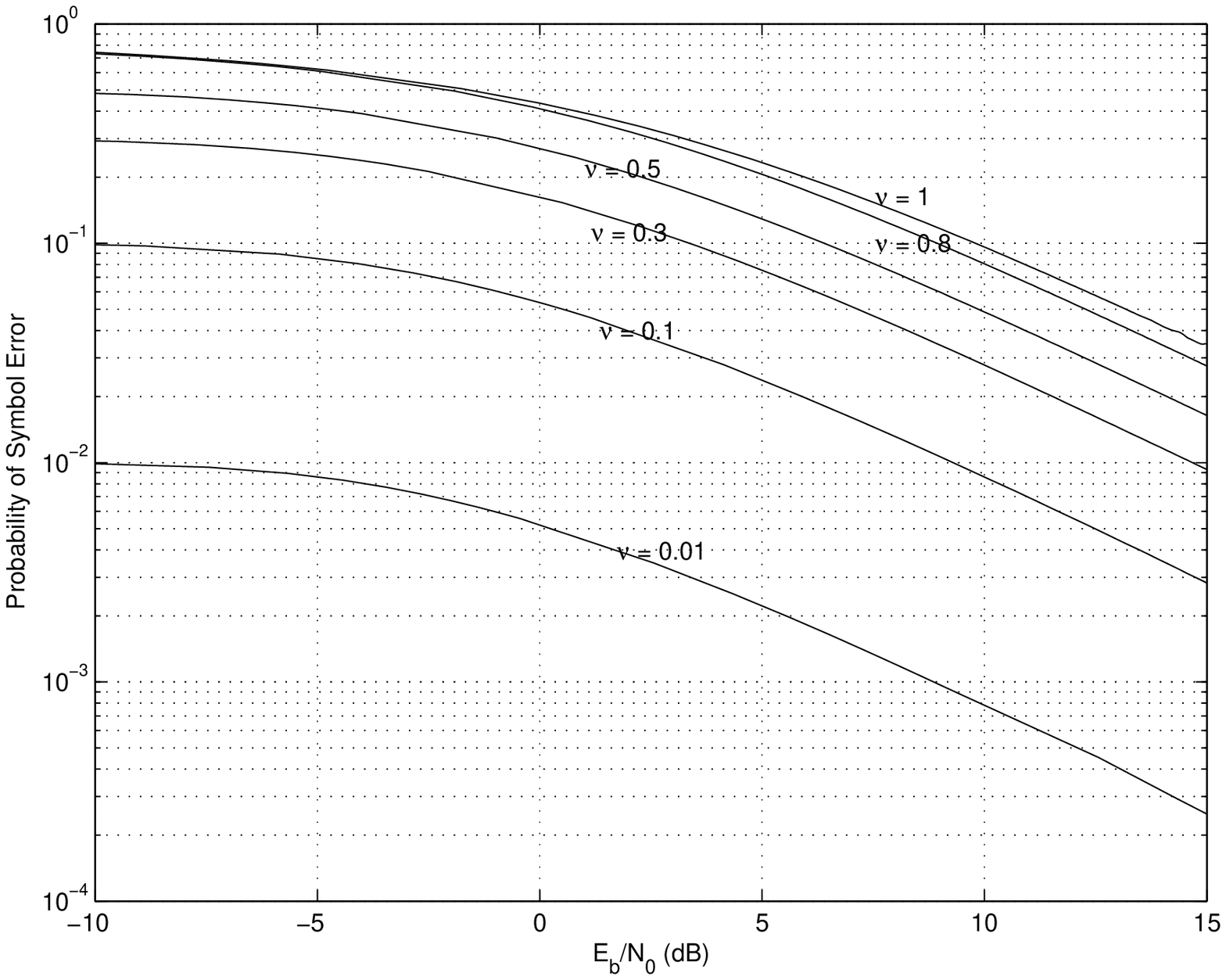}
\caption{Probability of symbol error vs. $E_b/N_0$ for 8-OOPSK
signaling with different duty factor values, $\nu$, in coherent
Rayleigh fading channels with $E\{|h|^2\} = 1$.} \label{fig:oopskM8}
\end{center}
\end{figure}

Figure \ref{fig:oopskM4} plots the average error probability curves
for 4-OOPSK signaling with different duty cycle values in the
coherent Rayleigh fading channel with $E\{|h|^2\} = 1$. In ordinary
$M$-PSK modulation, each symbol carries $\log_2 M$ bits. It is
important to note that in $M$-OOPSK modulation, the maximum number
of bits that can be carried by each symbol is equal to the entropy $
H(\nu) = \nu \log_2 (M/\nu) + (1 - \nu) \log_2 (1 / (1 - \nu)) $
which decreases to zero as $\nu \to 0$. Hence, decreasing the duty
cycle diminishes the data rates. For fair comparison, Fig.
\ref{fig:oopskM4} plots the curves as a function of the SNR
normalized by the entropy of the $M$-OOPSK source, giving the SNR
per bit. It is observed from the figure that if the peakedness of
the input signals is increased sufficiently (e.g., $\nu = 0.3, 0.1,
0.01$), significant improvements in error performance are achieved
over the ordinary PSK (i.e., OOPSK with $\nu =1$) performance. This
is due to the fact that the minimum distance of the constellation
increases with decreasing $\nu$. We note that 4-OOPSK with $\nu =
0.8$ performs worse than regular 4-PSK. As expected from Remark
\ref{rem:lowasymptotics}, for $\nu < 0.8$, error probabilities
approach $\nu$ as $\tsnr \to 0$. Fig. \ref{fig:oopskM8} plots the
average error probabilities for 8-OOPSK signaling again as a
function of the SNR per bit in the coherent Rayleigh fading channel
with $E\{|h|^2\} = 1$. Similarly, it is observed that the error
performance improves with decreasing duty cycle.
%if $\nu
%\le 0.5$ and significant gains are obtained when $\nu = 0.1$ which
%reguires a 10-fold increase in the peak-to-average power ratio when
%compared to ordinary QPSK signaling. It is interesting to note that
%if $\nu < 0.8$, the threshold value $T_q \to \infty$ as $P\to 0$.
%Therefore, for sufficiently small $P$, the zero signal, $x_0$, is
%always chosen as the detected signal and the error probability
%becomes $\nu$. Indeed, it is observed in Fig.
%\ref{fig:ooqpskrayleigh} that the error curves corresponding to
%OOQPSK signaling with $\nu < 0.8$ approach to $\nu$ as SNR
%decreases. If $\nu > 0.8$, $T_q \to -\infty$ as $P \to 0$. When $T_q
%\le 0$, the zero signal is never detected. Note that this behavior
%is also exhibited if OOBPSK modulation is used.

\subsection{OOFSK} \label{sec:coherentfading}

In this section, we assume that OOFSK signals are transmitted. The
output of the bank of $M$ demodulators is $\y =
(y_1,y_2,\ldots,y_M)$. It is readily observed from
(\ref{eq:outputoofsk}) that conditioned on $h$ and $s_i$, $y_m$ is a
proper complex Gaussian random variable with mean
$E\{y_m|s_i,h\}=\alpha he^{j\theta_i}\delta_{mi}$ and variance
$var\{Y_m|s_i,h\}=E{|n|^2}=1$ where $\delta_{mi} = 1$ if $m = i$ and
0 otherwise. We assume that energy detection is employed. Hence, the
detector observes $\R = (R_1, R_2, \ldots, R_M)$ where $R_m =
|y_m|^2$ which gives the energy in the $m^{\text{th}}$ frequency.
Conditioned on $s_i$ and $|h|$, $R_m = |y_m|^2$ is a chi-square
random variable with two degrees of freedom and the conditional
probability density function of $R_m$ is given by
\begin{eqnarray}
f_{R_m|s_i,|h|}(R_m|s_i,|h|)=\left\{\begin{array}{ll}e^{-(\alpha^2|h|^2+R_m)}I_0\left(2\sqrt{\alpha^2|h|^2
R_m}\right) \!\!\!&  m=i\\
e^{-R_m} \!\!\!& m\neq i \end{array}\right. \label{eq:chisquarepdf}
\end{eqnarray}
where $m \in \{1,\ldots,M\}$ and $i \in \{0,1,\ldots,M\}$. Note that
$\{R_i\}$ are i.i.d. random variables. The conditional joint
distribution function of $\R$ is given by
\begin{gather}\label{pdf:Rician}
f_{\R|s_i, |h|}(\R|s_i,|h|)=\left\{\begin{array}{ll}
e^{-\sum_{j=1}^MR_j}e^{-\alpha^2|h|^2}I_0\left(2\sqrt{R_i\alpha^2|h|^2}\right)
& 1\leq i\leq M\\
e^{-\sum_{j=1}^MR_j}& i=0
\end{array}\right.
\end{gather}
Again, MAP decision rule is used for the detection of the OOFSK
signals. Hence, $s_i(t)$ is the detected signal if the following
condition is satisfied:
\begin{equation}\label{eq:mapfsk}
p_i f_{\R|s_i, |h|}(\R| s_i , |h|) > p_j f_{\R|s_j, |h|}(\R | s_j ,
|h|) \quad \forall j \neq i
\end{equation}
Note that we again have $p_i = \frac{\nu}{M}$ for $i \neq 0$ and
$p_0 = 1-\nu$. The following proposition provides a simplified MAP
decision rule and a closed-form expression for the error
probability.
\begin{prop:oofskc} \label{prop:oofskc}
The optimal MAP decision rule for OOFSK signaling in coherent fading
channels is in the following form: $s_i(t)$ for $i \neq 0$ is
detected if
\begin{align}\label{eq:smapfsk}
R_i > R_j \  \forall j \neq i \quad \text{and} \quad R_i > \tau =
\left\{
\begin{array}{ll}
\frac{\left[I_0^{-1}\left(\xi\right)\right]^2}{4\alpha^2|h|^2} & \xi
\ge 1
\\
0 & \xi < 1
\end{array}\right.,
\end{align}
where $ \xi=\frac{M(1-\nu)\,e^{\alpha^2|h|^2}}{\nu}$ and $I_0^{-1}$
is the functional inverse of the zeroth order modified Bessel
function of the first kind. No transmission and hence $s_0(t)$ is
detected if $R_i < \tau \,\,\, \forall i.$ The error probability of
OOFSK modulation as a function of the instantaneous realization of
the fading magnitude is
\begin{gather}
P_{e||h|}=1- ((1-\nu) P_{c|s_0,|h|}+\nu P_{c|s_1,|h|})
\label{eq:Peoofsk}
\end{gather}
where \vspace{-1cm}
\begin{gather}
P_{c|s_0,|h|} = (1-e^{-\tau})^M, \text{ and }
\end{gather}
\begin{gather}
P_{c|s_1,|h|} =
\sum_{n=0}^{M-1}\frac{(-1)^n}{n+1}\left(\begin{array}{cc}M-1\\n\end{array}\right)
e^{-\frac{n}{n+1}\alpha^2|h|^2}
Q_1\left(\sqrt{\frac{2}{n+1}}\alpha|h|,\sqrt{2(n+1)\tau}\right),
\label{eq:Pc1oofsk}
\end{gather}
are the correct detection probabilities when $s_0(t)$ and $s_1(t)$,
respectively, are the transmitted signals. In the above formulation,
$Q_1(\cdot,\cdot)$ denotes the Marcum $Q$-function \cite{Simon}.
\end{prop:oofskc}

\emph{Proof}: Using the prior probabilities and the conditional
joint distribution function in (\ref{pdf:Rician}), the detection
rule in (\ref{eq:mapfsk}) can easily be simplified to that in
(\ref{eq:smapfsk}) by noting that $I_0(x)$ is a monotonically
increasing function of $x \ge 0$. Moreover, due to this monotonicity
and the fact that $I_0(0) = 1$, the inverse function $I_0^{-1}(x)$
is well-defined for $x \ge 1$.

Having obtained the decision rules, we first express the correct
detection probability conditioned on $s_1(t)$ being the transmitted
signal and $|h|$:
\begin{align}\small\label{PE_OOFSK1}
P_{c|s_1, |h|}&= P(R_1> R_2,\ldots,R_1>R_M, R_1 > \tau \,\,|\,\,
s_1, |h|)
\\
&=\int_{\tau}^\infty
P(R_1>R_2,\ldots,R_1>R_M\,\,|\,\,R_1=x,s_1,|h|)f_{R_1|s_1,|h|}(x|s_1,|h|)dx
\\
&=\int_{\tau}^\infty(1-e^{-x})^{M-1}e^{-(\alpha^2|h|^2+x)}I_0\left(2\sqrt{\alpha^2|h|^2x}\right)dx
\label{eq:Pcoofsks11}
\\
&=\sum_{n=0}^{M-1}(-1)^n
\left(\begin{array}{cc}M-1\\n\end{array}\right)\int_{\tau}^\infty
e^{-nx}e^{-(\alpha^2|h|^2+x)}I_0\left(2\sqrt{\alpha^2|h|^2x}\right)dx
\label{eq:Pcoofsks12}
\end{align}
where (\ref{eq:Pcoofsks11}) follows by noting from
(\ref{eq:chisquarepdf}) that $\{R_i\}_{i=2}^M$ are independent and
identically distributed exponential random variables given that
$s_1(t)$ is sent, and (\ref{eq:Pcoofsks12}) is obtained by using the
binomial expansion $(1 - e^{-x})^{M - 1} = \sum_{n=0}^{M-1}
\left(\begin{array}{cc}M-1\vspace{-.3cm}\\n\end{array}\right) (-1)^n
e^{-nx}. $ The Marcum Q-function is defined as
$Q_1(\alpha,\beta)=\int_{\beta}^\infty
xe^{-\frac{x^2+\alpha^2}{2}}I_0(\alpha x)dx$ \cite{Simon}. By
applying a change of variables, the integral in
(\ref{eq:Pcoofsks12}) can be expressed as a Marcum $Q$-function,
leading to the expression in (\ref{eq:Pc1oofsk}). Note that since
the FSK signals are orthogonal, the correct detection probabilities
for nonzero signals other than $s_1(t)$ are also equal to
(\ref{eq:Pc1oofsk}). When $s_0(t)$ is transmitted, the correct
detection probability is
\begin{align}
P_{c| s_0, |h|}&=P(R_1<\tau,,\ldots,R_M<\tau \,\,|\,\, s_0, |h|) =
\left(\int_0^{\tau} e^{-x}dx\right)^M = (1-e^{-\tau})^M.
\end{align}
Finally, the error probability can be written as in
(\ref{eq:Peoofsk}). \hfill $\square$
\begin{rem:avgerrorproboofsk}
The probability of error averaged over the instantaneous
realizations of the fading magnitude is obtained from
$P_e=\int_{0}^{\infty}P_{e||h|} \, dF_{|h|}|h|$ where $F_{|h|}$ is
the distribution function of the fading magnitude $|h|$.
\end{rem:avgerrorproboofsk}

%When $v=1$, OOFSK signaling reduces to ordinary FSK. In this
%case, (\ref{OOFSKrule}) becomes:
%\begin{equation}  \label{FSKrule1}
%\frac{1}{M} e^{-\alpha^2|h|^2}I_0(2\sqrt{R_k\alpha^2|h|^2})>0.
%\end{equation}
%It is easily seen that (\ref{FSKrule1}) is always satisfied. So,
%(\ref{FSKrule}) is enough for the detection of FSK signaling. Now,
%assuming signal $s_1$ is transmitted, the probability of error is:
%\begin{align}\tiny \label{Pee}
%P_e&=1-P(R_2<R_1,\ldots,R_M<R_1 | s_1)\nonumber\\
%&=1\!\!-\!\!\!\int_0^\infty \!\!\!\!\!P(R_2\!<\!R_1,\ldots,R_M\!\!<\!R_1|R_1=x)f_{R_1|s_1}(x | s_1)dx\nonumber\\
%&=-\sum_{n=1}^{M-1}(-1)^n\left(\begin{array}{cc}M-1\\n\end{array}\right)\frac{1}{n+1}e^{-\frac{n}{n+1}\alpha^2|h|^2}.
%\end{align}
%Due to the symmetry of the channel, when signals other than $s_1$
%are transmitted, we have the same probability of error. Hence,
%(\ref{Pee}) is the probability of error for FSK with $|h|$ known to
%the receiver. Note that since $Q_1(\cdot, 0) = 1$, if we choose $T_h
%= 0$, (\ref{eq:oofskerror}) becomes equal to (\ref{Pee}). We obtain
%the average probability of error $\bar{P}_e$:
%\begin{align}\tiny
%\bar{P}_e&=\int_0^\infty P_e f_{|h|^2}(x)dx\nonumber\\
%&=\int_0^\infty
%P_e\frac{1}{\gamma^2}e^{-\frac{|d|^2+x}{\gamma^2}}I_0\left(\frac{2|d|\sqrt{x}}{\gamma^2}\right)dx\nonumber\\
%&=-\sum_{n=1}^{M-1}(-1)^n\left(\begin{array}{cc}
%M-1\\n\end{array}\right)\frac{1}{(n+1)\gamma^2}e^{-\frac{|d|^2}{\gamma^2}}\nonumber\\
%&\quad\times\int_0^\infty
%e^{-\frac{n\alpha^2|x|^2}{n+1}}e^{-\frac{x}{\gamma^2}}I_0\left(\frac{2|d|\sqrt{x}}{\gamma^2}\right)dx.
%\end{align}
\begin{rem:oofsk}
Conclusions similar to those in Remarks \ref{rem:lowasymptotics} and
\ref{rem:highasymptotics} in Section \ref{subsec:coopsk} can be
drawn in this section as well. We again have
\begin{gather}
\lim_{\tsnr \to 0}\tau =\left\{
\begin{array}{ll}
\infty & \nu < \frac{M}{M+1}
\\
0 & \nu \ge \frac{M}{M+1}
\end{array}\right. \quad \text{and} \quad \lim_{\tsnr \to \infty}\tau =
\infty. \label{eq:asymptoticsoofsk}
\end{gather}
The second limit in (\ref{eq:asymptoticsoofsk}) can easily be shown
by noting the fact that $I_0^{-1}(x) \ge \log x $. Using this fact,
we have $\tau =
\frac{\left[I_0^{-1}\left(\xi\right)\right]^2}{4\alpha^2|h|^2} \ge
\frac{\log^2 \xi }{4\alpha^2|h|^2} = \frac{\left(\alpha^2 |h|^2 +
\log \frac{M(1-\nu)}{\nu}\right)^2}{4\alpha^2|h|^2}\to \infty$ as
$\alpha \to \infty$. As a result, the correct detection
probabilities show the same behavior as those in Remarks
\ref{rem:lowasymptotics} and \ref{rem:highasymptotics}.
\end{rem:oofsk}

\begin{figure}
\begin{center}
\includegraphics[width = 0.65\textwidth]{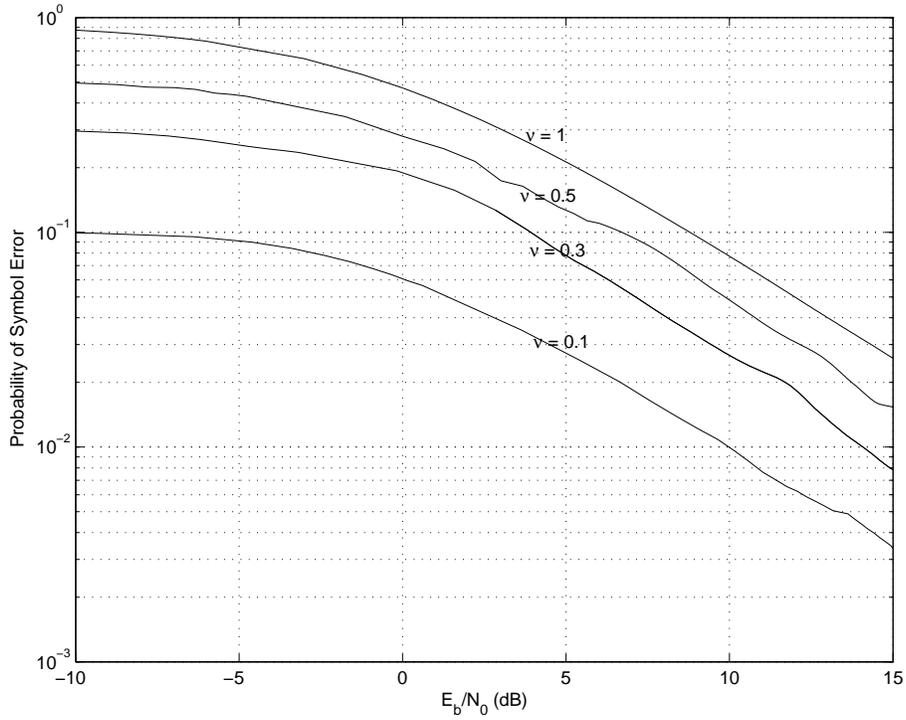}
\caption{Probability of error of 16-OOFSK in the coherent Rayleigh
fading channel with $E\{|h|^2\} = 1$. The duty factor values are
$\nu = 1,,0.5, 0.3$ and $0.1$.} \label{fig:oofskM16}
\end{center}
\end{figure}

Figure \ref{fig:oofskM16} plots the average probability of error
values of 16-OOFSK as a function of $\tsnr$ per bit for different
values of duty cycle parameter $\nu$ in the Rayleigh fading channel
with $E\{|h|^2\} = 1$. Similarly as in Section \ref{subsec:coopsk},
OOFSK signaling with low duty cycle has superior performance in
terms of error rates. From another perspective, if the duty cycle of
the modulation is reduced, the same performance can be achieved at
smaller $\tsnr$ per bit values, improving the energy efficiency.

%Setting $h=1$ in (\ref{pdf:Rician}), (\ref{FSKrule}),
%(\ref{OOFSKrule}) and (\ref{PE_OOFSK}), we get the joint
%distribution function, decision rules and probability of error for
%OOFSK signaling over the unfaded Gaussian channel. The probability
%of error of FSK signaling over the Gaussian channel \cite{Proakis}
%is:
%\begin{eqnarray}
%P_e=\sum_{n=1}^{M-1}(-1)^{n+1}\left(\begin{array}{cc}M-1\\n\end{array}\right)\frac{1}{n+1}e^{-\frac{n\rho}{n+1}}.
%\end{eqnarray}
%where $\rho$ is the SNR/symbol.\\ Fig. \ref{Gaussian} is the
%comparison of FSK and OOFSK in the unfaded Gaussian channel.
%Similarly, we observe that decreasing the duty cycle, $v$, improves
%the energy efficiency.

\section{Error Probability in Noncoherent Fading Channels}
\label{sec:noncoherent}

In this section, we consider the scenario in which neither the
transmitter nor the receiver knows the instantaneous realizations of
the fading coefficients. We consider a fast Rician fading
environment and hence $\{h_k\}$ is a sequence of i.i.d. proper
complex Gaussian random variables with mean $E\{h_k\} = d$ and
variance $E\{|h_k|^2\} = \gamma^2$. It is assumed that channel
statistics and hence the values of $d$ and $\gamma^2$ are assumed to
be known at the transmitter and receiver. Note that this model also
represents scenarios in which training symbols are employed to
facilitate channel estimation at the receiver, and the transmitter
interleaves the data symbols for protection against error bursts. In
such cases, $d$ and $\gamma^2$ represent the channel estimate and
the estimate error, respectively\footnote{It should be noted that in
training-based schemes, $d$ and $\gamma^2$ are dependent on
$\tsnr$.}. Moreover, due to interleaving, the data symbols
experience independent fading.

\subsection{OOPSK}

Similarly as in Section \ref{sec:coherent}, MAP detection is
employed at the detector. Hence, $s_i$ is the detected signal if
\vspace{-.7cm}
\begin{gather}\label{eq:mapn}
p_i f_{y|s_i} (y | s_i) > p_j f_{y|s_j}(y| s_j) \quad \forall j \neq
i
\end{gather}
where the conditional probability density function in the absence of
receiver channel knowledge is now given by \vspace{-.5cm}
\begin{gather} \label{eq:cpdfn}
f_{y|s_i}(y|s_i) = \left\{
\begin{array}{ll}
\frac{1}{\pi (\alpha^2 \gamma^2  + 1)}\, e^{-\frac{|y - \alpha d
e^{j\theta_i}|^2}{\alpha^2 \gamma^2  + 1}} & 1 \le i \le M \\
\frac{1}{\pi} \, e^{-|y|^2} & i=0
\end{array}\right..
\end{gather}
%In  this section, we assume that $|d|
%> 0$ because phase-shift keying cannot be employed to carry information
%over an unknown Rayleigh fading channel.
The following proposition describes the optimal decision regions and
provides an expression for the error probability of OOPSK signaling
in noncoherent Rician fading channels. Note that the results
immediately specialize to Rayleigh fading channels when it is
assumed that $d = 0$.
%Note further that in such a case, we have the
%performance of on-off keying because PSK cannot be used to carry
%information over unknown Rayleigh fading channels.
\begin{prop:oopsknc}
The optimal decision regions for OOPSK signals when transmitted over
noncoherent Rician fading channels are
\begin{gather}
\D_i = \left\{ y = |y|e^{j\theta_y}: \frac{2\pi (i-\frac{3}{2})}{M}
\le \theta_y \le \frac{2\pi (i-\frac{1}{2})}{M} \text{ and } |y|^2 +
\frac{2 |d|}{\alpha \gamma^2}\, |y| \cos(\theta_y - \theta_i)
> \tau\right\} \quad 1\le i \le M, \nonumber
\\
\D_0 = \left\{y = |y|e^{j\theta_y}: |y|^2 + \frac{2
|d|}{\alpha\gamma^2}\, |y| \cos(\theta_y - \theta_i) < \tau \quad
\forall i \neq 0\right\} \label{eq:droopskn}
\end{gather}
where $\tau = \left\{
\begin{array}{ll}
\zeta & \zeta \ge 0
\\
0 & \zeta < 0
\end{array}\right.$ and $\zeta = \frac{|d|^2}{\gamma^2} + \left( 1 +
\frac{1}{\alpha^2\gamma^2 } \right) \ln\left( \frac{M(1-\nu)}{\nu}
\left( 1+ \alpha^2\gamma^2 \right)\right).$ Furthermore, the error
probability is given by
\begin{gather}
P_e = 1 - ((1-\nu)P_{c|s_0} + \nu P_{c|s_1}) \label{eq:Peoopskn}
\end{gather}
where \vspace{-.5cm}
\begin{align}
P_{c|s_0}&= M \!\!\int_0^\htau \!\!\left(1 - 2Q\left(\sqrt{2}x \tan
\frac{\pi}{M}\right)\right) \frac{1}{\sqrt{\pi}}\, e^{-x^2}\, dx
-M\!\! \int_\htau^\ttau \!\!\left(1 - 2Q\left(\sqrt{2} \sqrt{\tau -
x^2 - \frac{2|d|}{\alpha\gamma^2 }x}\right)\right)
\frac{1}{\sqrt{\pi}}\, e^{-x^2}\, dx, \label{eq:Pcoopskns0}
\\
&\qquad \quad \quad P_{c|s_1} =\int_{\htau}^\infty \left(1 -
2Q\left(\frac{\sqrt{2}\,x \tan\frac{\pi}{M}}{\sqrt{1 + \alpha^2
\gamma^2 }} \right)\right) \frac{1}{\sqrt{\pi(1 + \alpha^2\gamma^2
)}}\, e^{-\frac{(x-\alpha|d|)^2}{1+\alpha^2\gamma^2 }} \, dx
\nonumber
\\
&\,\,\,\,\,\,\qquad \quad \quad \qquad-\int_\htau^\ttau \left( 1 - 2
Q\left(\frac{\sqrt{2}\sqrt{\tau - x^2 - \frac{2|d|}{\alpha\gamma^2
}x}}{\sqrt{1 + \alpha^2\gamma^2 }} \right)\right)
\frac{1}{\sqrt{\pi(1 + \alpha^2\gamma^2 )}}\,
e^{-\frac{(x-\alpha|d|)^2}{1+\alpha^2\gamma^2 }} \, dx
\label{eq:Pcoopskn}
\end{align}
are the correct detection probabilities when $s_0(t)$ and $s_1(t)$,
respectively, are the transmitted signals. In the above integral
expressions, $\htau = \frac{1}{1 + \tan^2
\frac{\pi}{M}}\left(\sqrt{\frac{|d|^2}{\alpha^2 \gamma^4}+\tau
\left(1+\tan^2 \frac{\pi}{M}\right)}\,-\frac{|d|}{\alpha\gamma^2
}\right)$ and $\ttau = \sqrt{\frac{|d|^2}{\alpha^2\gamma^4
}+\tau}-\frac{|d|}{\alpha\gamma^2}$.
\end{prop:oopsknc}

\emph{Proof}: The decision rule (\ref{eq:mapn}) can easily be
simplified to yield the following rule: $s_i$ for $i \neq 0$ is the
detected signal if
\begin{align}
\Re(y e^{-j\theta_i}) > \Re(y e^{-j\theta_j}) \quad \forall j \neq
i, 0  \quad \text{ and } \quad |y|^2 + \frac{2 |d|}{\alpha\gamma^2
}\, \Re(y e^{-j\theta_i})
> \tau \label{eq:smapn2}
\end{align}
where $ \tau$ is defined in the proposition. The signal $s_0$ is the
detected signal if
\begin{gather} \label{smapn0}
|y|^2 + \frac{2 |d|}{\alpha \gamma^2}\, \Re(y e^{-j\theta_i}) < \tau
\quad \forall i \neq 0.
\end{gather}
Using the same arguments as in Section \ref{sec:coherent}, we obtain
the above decision rules by assuming without loss generality that $d
= |d|$. Otherwise, in all the decision rules, $y$ must be replaced
by $\tilde{y} = y e^{-\theta_d}$ where $\theta_d$ is the phase of
fading mean $d$, which is known at the receiver. These detection
rules lead to the decision regions given in (\ref{eq:droopskn}).

The correct detection probability of a nonzero signal can be found
by considering, without loss of generality, that $s_1(t)$ with phase
$\theta_1 = 0$ is sent. Given that $s_1(t)$ is sent, the correct
detection probability can be expressed as
\begin{align}
\!\!\!\!\!\!\!\!\!P_{c|s_1} \!&=\! P(y \in \D_1|s_1) = P\left(y_r
> \htau, |y_i| < y_r \tan \frac{\pi}{M}\,|\,s_1\right) \!-\!P\left(
\htau < y_r < \ttau, |y_i| < \sqrt{\tau - y_r^2 -
\frac{2|d|}{\gamma^2 \alpha}y_r}|s_1\right)
\label{eq:pcrroroopskns1}
\end{align}
where $y = y_r + j y_i$. We note that in the $(y_r, y_i)$ plane,
$\htau$ is the horizontal axis value of the point at which the line
$y_i = y_r \tan \frac{\pi}{M}$ and the circle $y_r^2 + y_i^2 +
\frac{2|d|}{\alpha \gamma^2}y_r = \tau$ intersect, and $\ttau$ is
the value of the point at which the same circle intersects the
horizontal axis. The correct detection probability for $s_0(t)$ is
\begin{align}
P_{c|s_0} &= MP\left( 0 < y_r < \htau, |y_i| < y_r \tan
\frac{\pi}{M} \, | \,s_0\right) + MP\left( \htau < y_r < \ttau,
|y_i| < \sqrt{\tau - y_r^2 - \frac{2|d|}{\alpha\gamma^2}y_r}\, |
\,s_0 \right). \label{eq:pcrroroopskns0}
\end{align}
The correct detection probability expressions in
(\ref{eq:Pcoopskns0}) and (\ref{eq:Pcoopskn}) are the integral
representations of (\ref{eq:pcrroroopskns0}) and
(\ref{eq:pcrroroopskns1}), respectively, obtained by representing
the inner integrals by $Q$-functions. Finally, the error probability
can be obtained from the correct detection probabilities as in
(\ref{eq:Peoopskn}).\hfill $\square$

\begin{rem:drsoopskn}
Note that $|y|^2 + \frac{2 |d|}{\alpha \gamma^2}\, |y| \cos(\theta_y
- \theta_i) < \tau$ defines a circular area. Therefore, $\D_0$ is
the intersection of $M$ circular regions.
\end{rem:drsoopskn}

\begin{rem:lowasymptoticsoopskn}
It can again be easily verified that
\begin{gather}
\lim_{\tsnr \to 0} \tau = \left\{
\begin{array}{ll}
\infty & \nu < \frac{M}{M+1}
\\
0 & \nu > \frac{M}{M+1}
\end{array}\right..
\end{gather}
Therefore, similarly as in Remark \ref{rem:lowasymptotics}, if $\nu
< \frac{M}{M+1}$, $\lim_{\tsnr \to 0} P_e = \nu$. On the other hand,
if $\nu > \frac{M}{M+1}$, decision region $\D_0$ vanishes with
decreasing $\tsnr$.
\end{rem:lowasymptoticsoopskn}

\begin{rem:highasymptoticsoopskn}
As $\tsnr \to \infty$, we immediately observe that $\tau \to
\infty$, and as a result, $\htau \to \infty$, and $\ttau \to
\infty$. Asymptotically, $\D_0$ expands and becomes the complex
plane. This leads us to conclude that $P_{c|s_0} \to 1$. We observe
a different behavior from $P_{c|s_1}$. As $\tsnr$ increases, the
second integral in the expression of $P_{c|s_1}$ in
(\ref{eq:Pcoopskn}) vanishes. After applying a change of variables
with $\hat{x} = (x-\alpha|d|)/\sqrt{1+\alpha^2\gamma^2}$, the first
integral in the expression of $P_{c|s_1}$in (\ref{eq:Pcoopskn}) can
be written as
\begin{gather}
\int_{\frac{\htau-\alpha|d|}{\sqrt{1+\alpha^2\gamma^2}}}^\infty
\left(1 - 2Q\left(\sqrt{2}\tan\frac{\pi}{M} \left( \hat{x} +
\frac{\alpha|d|}{\sqrt{1+\alpha^2 \gamma^2}}\right)\right)\right)
\frac{1}{\sqrt{\pi}}\, e^{-\hat{x}^2} \, d\hat{x}.
\label{eq:Pcoopsks1firstinteg}
\end{gather}
Taking the limit of ($\ref{eq:Pcoopsks1firstinteg}$) as $\tsnr \to
\infty$ and hence $\alpha \to \infty$, we obtain
\begin{gather}
\lim_{\tsnr \to \infty} P_{c|s_1} =
\int_{-\frac{|d|}{\gamma}}^\infty \left(1 -
2Q\left(\sqrt{2}\tan\frac{\pi}{M} \left( \hat{x} +
\frac{|d|}{\gamma}\right)\right)\right) \frac{1}{\sqrt{\pi}}\,
e^{-\hat{x}^2} \, d\hat{x} \stackrel{\text{def}}{=}
P_{c,\infty|s_1}.
\end{gather}
Note that $P_{c|s_1}$ does not approach to 1 as $\tsnr \to \infty$.
Hence, we experience an error floor in noncoherent OOPSK signaling.
The error floor is due to the presence of the multiplicative noise
as a result of unknown fading. Even if the additive noise vanishes,
errors are possible due to the distortion caused by unknown random
phase shifts of fading. Note that $P_{c,\infty|s_1}$ depends only on
the Rician factor $\K = \frac{|d|^2}{\gamma^2}$ and $M$. If $\gamma
= 0$, then $\K = \infty$, and we have the unfaded Gaussian channel
for which we can easily see that $P_{c,\infty|s_1} = 1$. Hence, no
error floors exit in this case. If, on the other hand, $|d| = 0$ and
hence $\K = 0$, we have the unknown Rayleigh fading channel for
which $P_{c,\infty|s_1} = \frac{1}{M}$. Finally, as $\tsnr \to
\infty$, we have $P_e \to \nu (1 - P_{c,\infty|s_1})$. Therefore,
error floor decreases with decreasing duty cycle.
\end{rem:highasymptoticsoopskn}

\begin{figure}
\begin{center}
\includegraphics[width = 0.65\textwidth]{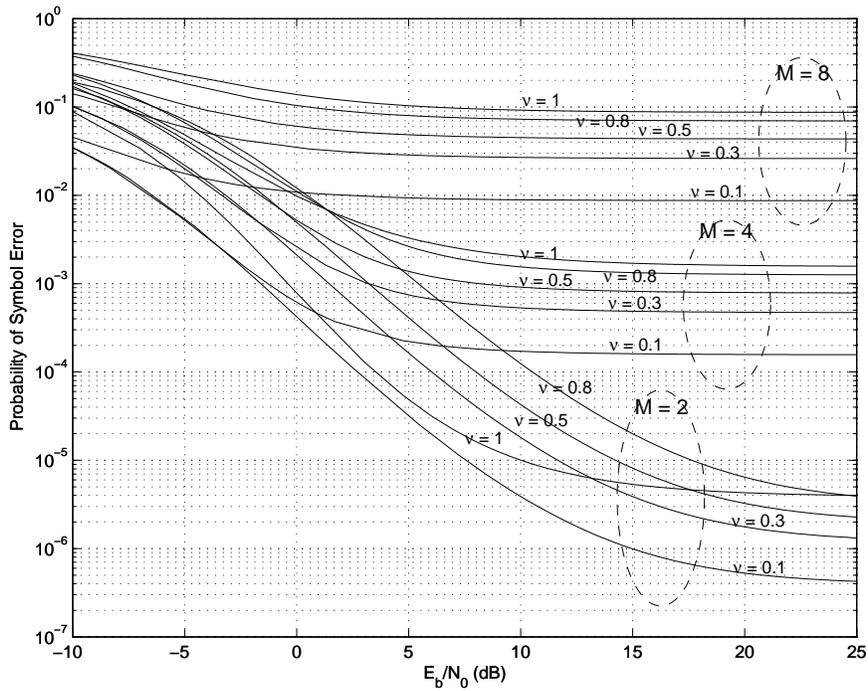}
\caption{Probability of symbol error vs. $E_b/N_0$ for OOPSK
signaling with duty factor values $\nu = 1,0.8,0.5,0.3,0.1$ and
constellation sizes $M = 2,4,8$ in the noncoherent Rician fading
channel with Rician factor $\K = |d|^2 / \gamma^2 = 10$.}
\label{fig:oopskn}
\end{center}
\end{figure}

Fig. \ref{fig:oopskn} plots the error probability curves for OOPSK
signaling as a function of SNR per bit in the noncoherent Rician
fading channel with Rician factor $\K = |d|^2/\gamma^2 = 10$. The
plots are provided for constellation sizes of $M = 2,4,$ and 8.
Similarly as before, it is seen in all cases that the error
performance improves as duty factor value decreases sufficiently.
Note that for 8-OOPSK, even having a duty value of $\nu = 0.8$
improves the performance with respect to the regular 8-PSK in the
entire range of SNR per bit values considered in the graph. On the
other hand, when $M = 2$, decreasing the duty cycle to $\nu = 0.8,
0.5$ or 0.3 does not provide gains with respect to the case of $\nu
= 1$ unless $\tsnr$ is high enough. As predicted, we observe error
floors in all cases. We note that error floors decrease with
decreasing constellation sizes and duty factors.

%It is again noted that for $\nu < 0.8$, $C_q \to \infty$ as $P \to
%0$. Hence, for sufficiently small $P$, the zero signal is always
%declared as the detected signal and the error probability becomes
%equal to $\nu$. Another interesting observation in Fig.
%\ref{fig:ooqpskriciannk5} is the existence of error floors for
%sufficiently high values of SNR. This is due to the fact that even
%if the additive noise vanishes, the performance is limited by the
%multiplicative noise introduced by unknown fading. Note that in the
%correct detection probability expressions (\ref{eq:Peqn1}) and
%(\ref{eq:Peqn1T0}), the arguments of the Q-function have the term
%$P$ in both the numerator and denominator. Hence, letting $P \to
%\infty$ or $N_0 \to 0$ does not lead the correct detection
%probabilities to 1.

\subsection{OOFSK}

In this section, we consider OOFSK signaling. If $s_i(t)$ is
transmitted, then $y_m$ is a proper complex Gaussian random variable
with mean $E\{y_m | s_i\}=\alpha d  e^{j\theta_i} \delta_{mi}$ and
variance $var\{y_m|s_i\}=\alpha^2\gamma^2 \delta_{mi}+1$ where
$\delta_{mi} = 1$ if $m=i$ and zero otherwise. Since $y_m$ is a
complex Gaussian random variable, $R_m = |y_m|^2$ is chi-square
distributed and the joint distribution function of the output vector
$\R$ conditioned on $s_i(t)$ being transmitted is
\begin{eqnarray}\label{Rician:unknown}
f_{\R|s_i}(\R|s_i)=\left\{\begin{array}{ll}
e^{-\sum_{\substack{j=1\\j\neq i}}^{M}R_j}
\frac{e^{-\frac{R_i+\alpha^2|d|^2}{1+\alpha^2\gamma^2}}}{1+\alpha^2\gamma^2}I_0\left(\frac{2\sqrt{R_i\alpha^2|d|^2}}{1+\alpha^2\gamma^2}\right)
& 1 \leq i\leq M\\
 e^{-\sum_{j=1}^MR_j}, & i=0
\end{array} \right..
\end{eqnarray}
The following result provides the optimal detection rule and the
error probability of OOFSK signaling in noncoherent Rician fading
channels.
\begin{prop:oofsknc}
The optimal MAP detection rule for OOFSK signaling over noncoherent
Rician fading channels is given as follows: $s_i(t)$ for $i\neq 0$
is the detected signal if
\begin{align} \label{FSK}
R_i > R_j \quad \forall j \neq i \quad \text{and} \quad R_i > \tau
=\left\{
\begin{array}{ll}
 \Phi^{-1}(\xi) & \xi \ge 1
 \\
 0 & \xi < 1
\end{array}\right.
\end{align}
where \vspace{-.5cm}
\begin{gather}
\Phi(x) = e^{\frac{\alpha^2 \gamma^2 x}{1 + \alpha^2 \gamma^2}}
I_0\left( \frac{2\sqrt{x \, \alpha^2 |d|^2}}{1 + \alpha^2
\gamma^2}\right) \text{ and } \xi = \frac{M(1-\nu)}{\nu} (1+
\alpha^2 \gamma^2) \, e^{\frac{\alpha^2 |d|^2}{1 + \alpha^2
\gamma^2}}.
\end{gather}
$s_0(t)$ is the detected signal if $R_i < \tau$ $\forall i$. The
probability of error is
\begin{equation}\label{eq:Peoofskn}
P_e=1 - ((1-\nu)P_{c|s_0} + \nu P_{c|s_1})
\end{equation}
where \vspace{-0.8cm}
\begin{gather}
P_{c|s_0} = \left(1-e^{-\tau}\right)^M \quad \text{ and},
\\
\!\!\!\!\!\!\!\!P_{c|s_1} \!=\!
\sum_{n=0}^{M-1}(-1)^n\!\!\left(\!\begin{array}{cc}M-1\\n\end{array}\!\right)
\frac{e^{-\frac{n\alpha^2|d|^2}{n(1+\alpha^2\gamma^2)+1}}}{n(1+\alpha^2\gamma^2)\!+\!1}
Q_1\!\!\left(\!\sqrt{\frac{2\alpha^2|d|^2}{(1\!\!+\!\alpha^2\gamma^2)[n(1\!\!+\!\!\alpha^2\gamma^2)\!\!+\!\!1]}},
\!\sqrt{\frac{2[n(1\!\!+\!\!\alpha^2\gamma^2)\!\!+\!\!1]\tau}{(1\!\!+\!\!\alpha^2\gamma^2)}}\right)
\label{eq:Pcoofskncs1}
\end{gather}
are the correct detection probabilities when $s_0(t)$ and $s_1(t)$,
respectively, are the transmitted signals.
\end{prop:oofsknc}
\emph{Proof}: The MAP decision rule in (\ref{FSK}) can be easily
obtained by using the conditional density function expression in
(\ref{Rician:unknown}) in the general MAP detection rule $p_i
f_{\R|s_i}(\R|s_i) > p_j f_{\R|s_j}(\R|s_j)$ for all $j \neq i$.
Since $\Phi$ is a monotonically increasing function and $\Phi(0) =
1$,
the functional inverse $\Phi^{-1}(x)$ is well-defined for $x \ge 1$. %In FSK
%signaling, the detection rule simplifies to $R_k>R_l, \forall l \neq
%k$, and the probability of error is readily found as
%\begin{align}  \label{PEfsk}
%P_e&=1-\int_0^\infty p_{R_k}(x)\left[\int_0^x p_{R_m}(y)dy\right]^{M-1}dx\\
%&=\!\!\!\sum_{n=1}^{M-1}(-1)^{n+1}\!\!\left(\!\!\!\begin{array}{cc}M-1\\n\end{array}\!\!\!\right)
%\frac{1}{n+1+n\alpha^2\gamma^2}e^{-\frac{n\alpha^2|d|^2}{n+1+n\alpha^2\gamma^2}}.
%\end{align}
%Not having a closed-form expression for $\Phi^{-1}$, we can find
%threshold value through numerical method according to the equation
%$T_h=\Phi^{-1}(T)$. Using this threshold value $T_h$, we can find a
%much more simplified expression for probability of error when there
%is signal is transmitted. For simplicity,
If $s_1(t)$ is assumed to be transmitted, then, similarly as in the
proof of Proposition \ref{prop:oofskc}, the probability of correct
detection is
\begin{align}
P_{c|s_1}&=P\left(R_1>R_2,\ldots,R_1>R_M, R_1>\tau \,\,|\,\, s_1\right)\\
&=\sum_{n=0}^{M-1}(-1)^n\left(\begin{array}{cc}M-1\\n\end{array}\right)
\int_{\tau}^\infty\frac{1}{1+\alpha^2\gamma^2}e^{-\frac{[n(1+\alpha^2\gamma^2)+1]x
+\alpha^2|d|^2}{1+\alpha^2\gamma^2}}I_0\left(\frac{2\alpha|d|\sqrt{x}}{1+\alpha^2\gamma^2}\right)dx.
\\
&=\sum_{n=0}^{M-1}(-1)^n\left(\begin{array}{cc}M-1\\n\end{array}\right)\frac{e^{-\frac{n\alpha^2|d|^2}{(n(1+\alpha^2\gamma^2)+1)}}}{n(1+\alpha^2\gamma^2)+1}
\int_{[n(1+\alpha^2\gamma^2)+1]\tau}^\infty\frac{e^{-\frac{x+\alpha^2|d|^2}{1+\alpha^2\gamma^2}}}{1+\alpha^2\gamma^2}I_0\left(\frac{2\alpha
|d|\sqrt{x}}{1+\alpha^2\gamma^2}\right)dx. \label{eq:subopterror}
\end{align}
The integral in (\ref{eq:subopterror}) can be expressed as a Marcum
$Q$-function and we obtain (\ref{eq:Pcoofskncs1}). When no signal is
transmitted, the probability of correct detection is
\begin{align}
P_{c|s_0}&=P\left(R_1<\tau,\ldots,R_M<\tau \,\, |\,\,
s_0\right)=\left(\int_0^{\tau}e^{-x}dx\right)^M=\left(1-e^{-\tau}\right)^M.
\end{align}
Finally, the probability of error is given as in
(\ref{eq:Peoofskn}). \hfill $\square$
\begin{rem:bounds}
The Marcum $Q$-function has the following bounds when $\beta
> \alpha>0$ \cite{Simon}:
\begin{eqnarray}
\frac{\beta}{\beta+\alpha}e^{-\frac{(\beta+\alpha)^2}{2}}\leq
Q_1(\alpha,\beta)\leq\frac{\beta}{\beta-\alpha}e^{-\frac{(\beta-\alpha)^2}{2}}.
\end{eqnarray}
Hence, in (\ref{eq:Pcoofskncs1}), when $\frac{\beta}{\alpha}=
\sqrt{\frac{[n(1+\alpha^2\gamma^2)+1]^2\tau}{\alpha^2|d|^2}}
> 1$, we can obtain bounds expressed using
exponential functions rather than the Marcum $Q$-functions.

If $|d| = 0$, we have the Rayleigh fading channel. In this case,
error probability expression simplifies immediately because
$Q_1(0,\beta) = e^{-\beta^2/2}$.
\end{rem:bounds}

\begin{rem:asymptoticsoofskn}
We can easily show that
\begin{gather}
\lim_{\tsnr \to 0}\tau =\left\{
\begin{array}{ll}
\infty & \nu < \frac{M}{M+1}
\\
0 & \nu > \frac{M}{M+1}
\end{array}\right. \quad \text{and} \quad \lim_{\tsnr \to \infty}\tau =
\infty. \label{eq:asymptoticsoofskn}
\end{gather}
From the first limit in (\ref{eq:asymptoticsoofskn}), we note that
if $\nu < \frac{M}{M+1}$, we again have $P_e \to \nu$ as $\tsnr \to
0$.

Let us now consider the second limit in
(\ref{eq:asymptoticsoofskn}). From (\ref{FSK}), we have $\Phi(\tau)
= \xi$ for large $\tsnr$. Using the fact that $I_{0}(x) \le e^{x}$
$\forall x\ge 0$, we have $\xi \le e^{\frac{\alpha^2 \gamma^2 \tau +
2\sqrt{\alpha^2 |d|^2 \tau}}{1+\alpha^2 \gamma^2}}$. By taking the
logarithm of both sides of this inequality, we can easily show the
second limit in (\ref{eq:asymptoticsoofskn}). Therefore, we can
easily see that $P_{c|s_0} \to 1$ as $\tsnr$ increases. Next, we
consider $P_{c|s_1}$. In the correct detection probability
expression in (\ref{eq:Pcoofskncs1}), the terms in the summation,
for which $n \neq 0$, approach to zero with increasing $\tsnr$ due
to the presence of $(n(1+\alpha^2 \gamma^2) + 1)$ in the denominator
and the fact that Marcum $Q$-function, being a probability, is upper
bounded by 1. When $n = 0$, the term in the summation is
$Q_1\left(\sqrt{\frac{2 \alpha^2
|d|^2}{1+\alpha^2\gamma^2}},\sqrt{\frac{2\tau}{1+\alpha^2\gamma^2}}\right)$.
Since $I_0(x) \ge 1$ $\forall x \ge 0$, we have $e^{\frac{\alpha^2
\gamma^2 \tau}{1+\alpha^2 \gamma^2}} \le \xi$ from which we can
easily observe, by taking the logarithm of both sides, that $\tau$
increases at most logarithmically with increasing $\alpha$.
Therefore, as $\alpha \to \infty$, $Q_1\left(\sqrt{\frac{2 \alpha^2
|d|^2}{1+\alpha^2\gamma^2}},\sqrt{\frac{2\tau}{1+\alpha^2\gamma^2}}\right)
\to Q_1\left(\sqrt{\frac{2|d|^2}{\gamma^2}},0\right) = 1$. Hence, as
$\tsnr \to \infty$, $P_{c|s_1} \to 1$, and hence $P_e \to 0$.
Therefore, unlike the OOPSK case, we do not have error floors in
this case due to the immunity of OOFSK with energy detection to
random phase rotations of fading.
\end{rem:asymptoticsoofskn}

\begin{figure}
\begin{center}
\includegraphics[width = 0.65\textwidth]{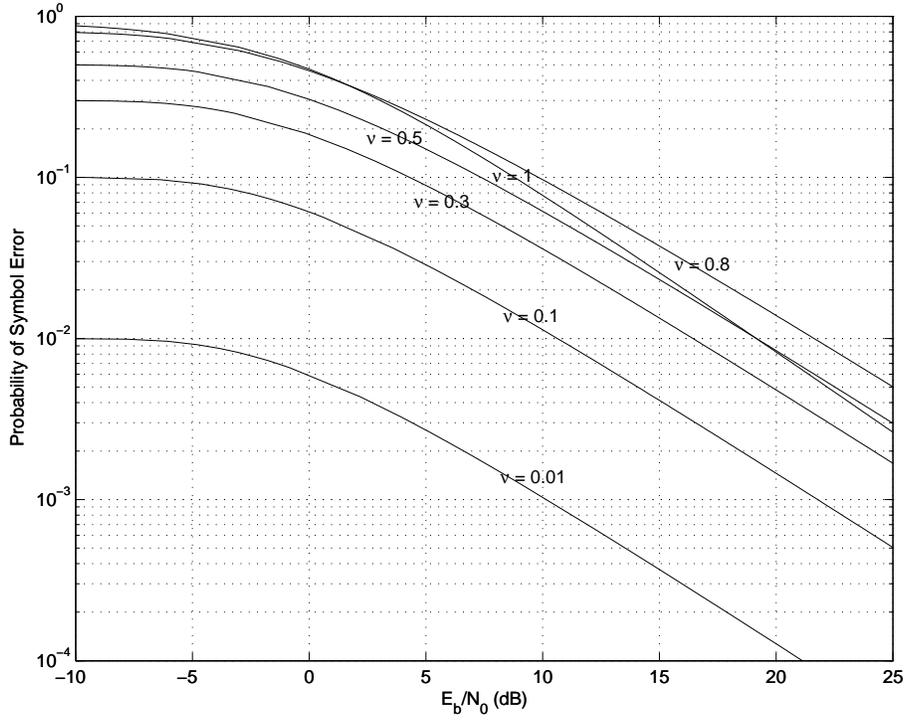}
\caption{Error probability of 16-OOFSK in the noncoherent Rayleigh
fading channel with duty factors $\nu=1,0.8,0.5, 0.3,0.1$, and
$0.01$.}\label{fig:oofsknK0}
\end{center}
\end{figure}

\begin{figure}
\begin{center}
\includegraphics[width = 0.65\textwidth]{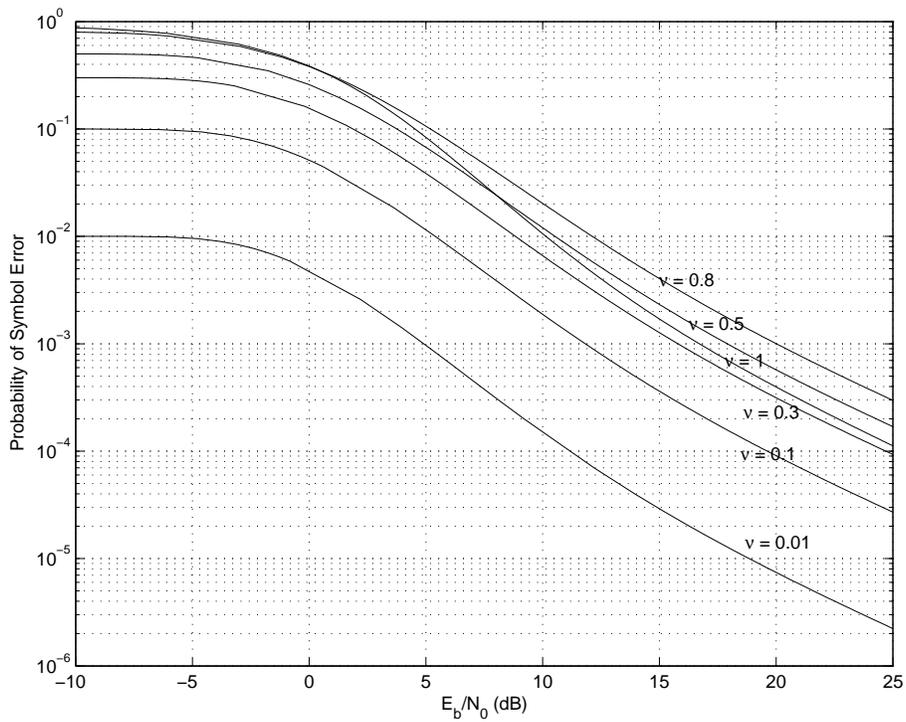}
\caption{Error probability of 16-OOFSK in the noncoherent Rician
fading channel with Rician factor $\K = \frac{|d|^2}{\gamma^2} = 5$.
The duty factor values are $\nu=1,0.8,0.5, 0.3,0.1$, and $0.01$.}
\label{fig:oofsknK5}
\end{center}
\end{figure}

Fig. \ref{fig:oofsknK0} provides the error rates of 16-OOFSK over
the noncoherent Rayleigh fading channel while Fig.
\ref{fig:oofsknK5} gives the error rates in the noncoherent Rician
channel with $\K = 5$. We observe that unlike the OOPSK case, OOFSK
performance is free of error floors at high $\tsnr$s. As $\tsnr$
decreases, we see that $P_e \to \nu$ for the cases in which $\nu <
M/(M+1)$. Moreover, we note that modulations with $\nu < 1$ perform
better than that with $\nu = 1$ at low $\tsnr$s. However, at high
$\tsnr$ levels, gains are realized if the duty factor is
sufficiently small.

%shows that the energy efficiency improves as one operates with lower
%duty cycle $v$. Fig. \ref{fig:bound} plots the simulated error
%probabilities and upper bounds for OOFSK modulation with $v  = 0.2
%\text{ and } 0.8$ when a suboptimum threshold value of $T_h =
%\text{SNR}/(\nu c)$ is used where $c$ is a parameter chosen to get
%the tightest upper bound.
%Fig. \ref{fig:Raleigh2} shows that the smaller the duty factor is,
%the more energy the OOFSK signaling saves.

\section{Random Coding Error Exponents of Peaky Signaling}
\label{sec:exponents}

In \cite{Gallager}, Gallager derived upper bounds on the probability
of error that can be achieved by block codes on general
discrete-time memoryless channels. Using an ensemble of codebooks
where each letter of each codeword is chosen independently of all
other letters with a certain probability distribution, it is shown
in \cite{Gallager} that for any rate $R$ less than the channel
capacity, the probability of error can be upper bounded by $ P_e \le
B e^{-N E(R)} $ where $B$ is a constant, $N$ is the codeword length,
and $E(R)$ is the random coding error exponent. $E(R)$ provides the
interactions between the probability of error, channel coding, data
rates, and the signal-to-noise ratio ({\snr}). The random coding
error exponent is obtained from
\begin{gather}\label{eq:E(R)}
E(R) = \sup_{0 \le \rho \le 1}  E_0(\rho) - \rho R
\end{gather}
where
\begin{gather} \label{eq:E0Fx}
E_0(\rho) = - \log \int_y \left( \frac{\nu}{M}\sum_{i = 1}^M f
_{y|s_i}(y|s_i)^{\frac{1}{1+\rho}} + (1-\nu)f
_{y|s_0}(y|s_0)^{\frac{1}{1+\rho}}\right)^{1+\rho} \ud y
\end{gather}
for OOPSK modulation and
\begin{gather}
E_0(\rho) = - \log \int_\R \left( \frac{\nu}{M}\sum_{i = 1}^M f
_{\R|s_i}(\R|s_i)^{\frac{1}{1+\rho}} + (1-\nu)f
_{\R|s_0}(\R|s_0)^{\frac{1}{1+\rho}}\right)^{1+\rho} \ud \R
\end{gather}
for OOFSK modulation. Note that for coherent fading channels,
$f_{y|s_i}$ and $f_{\R|s_i}$ are replaced by $f_{y|s_i,h}$ and
$f_{\R|s_i,|h|}$, respectively. In the coherent case, $E(R,h)$ is
also a function of $h$ and the average error exponent is obtained
from $E_h\{E(R,h)\}$ where $E_h$ denotes the expectation with
respect to $h$.

\begin{figure}
\begin{center}
\includegraphics[width = 0.65\textwidth]{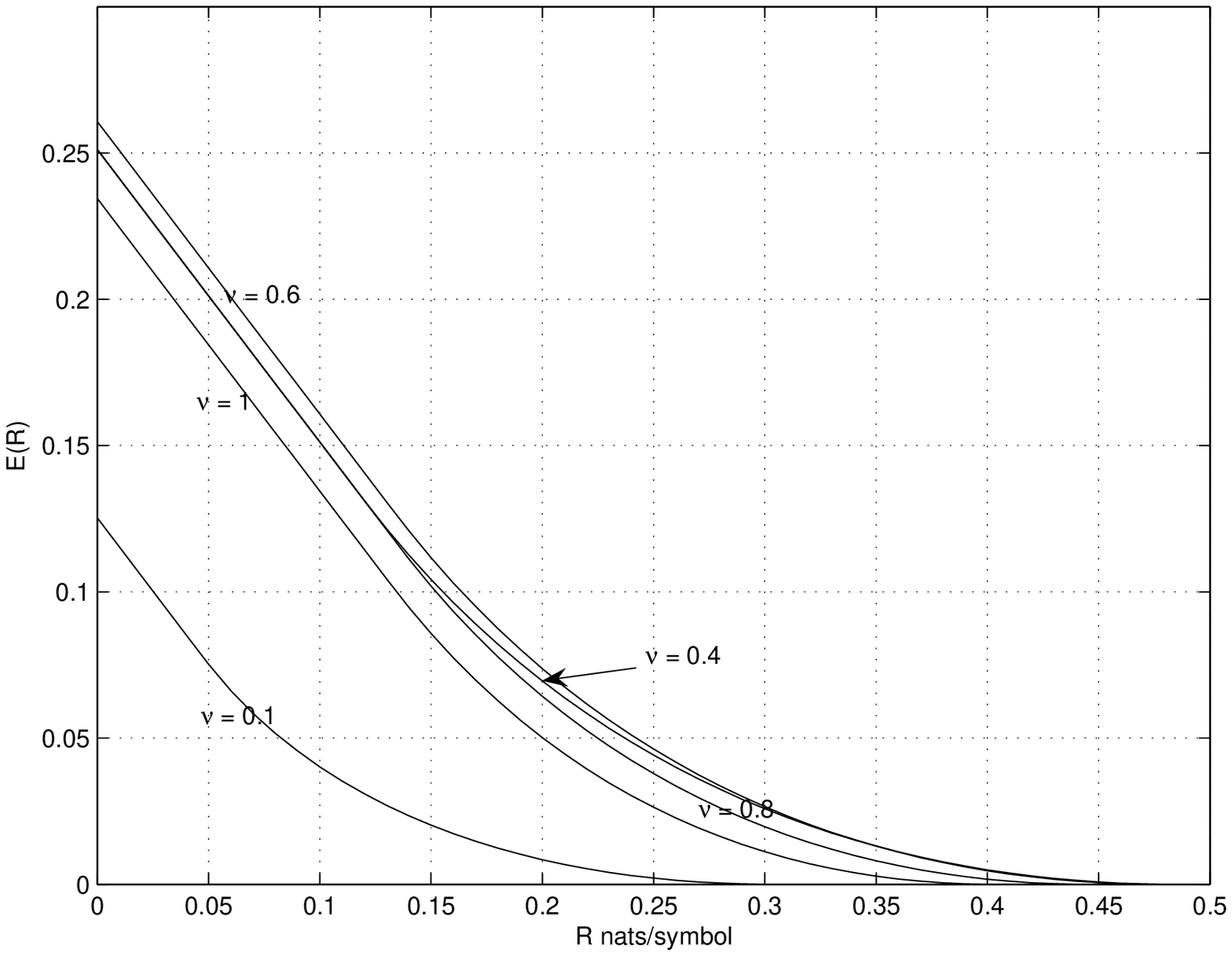}
\caption{Random coding error exponent of 16-OOPSK in the noncoherent
Rician fading channel with Rician factor $\K =
\frac{|d|^2}{\gamma^2} = 1$. The duty factor values are
$\nu=1,0.8,0.6, 0.4,0.1$. $\tsnr = 1$.} \label{fig:eeoopsknM16}
\end{center}
\end{figure}

\begin{figure}
\begin{center}
\includegraphics[width = 0.65\textwidth]{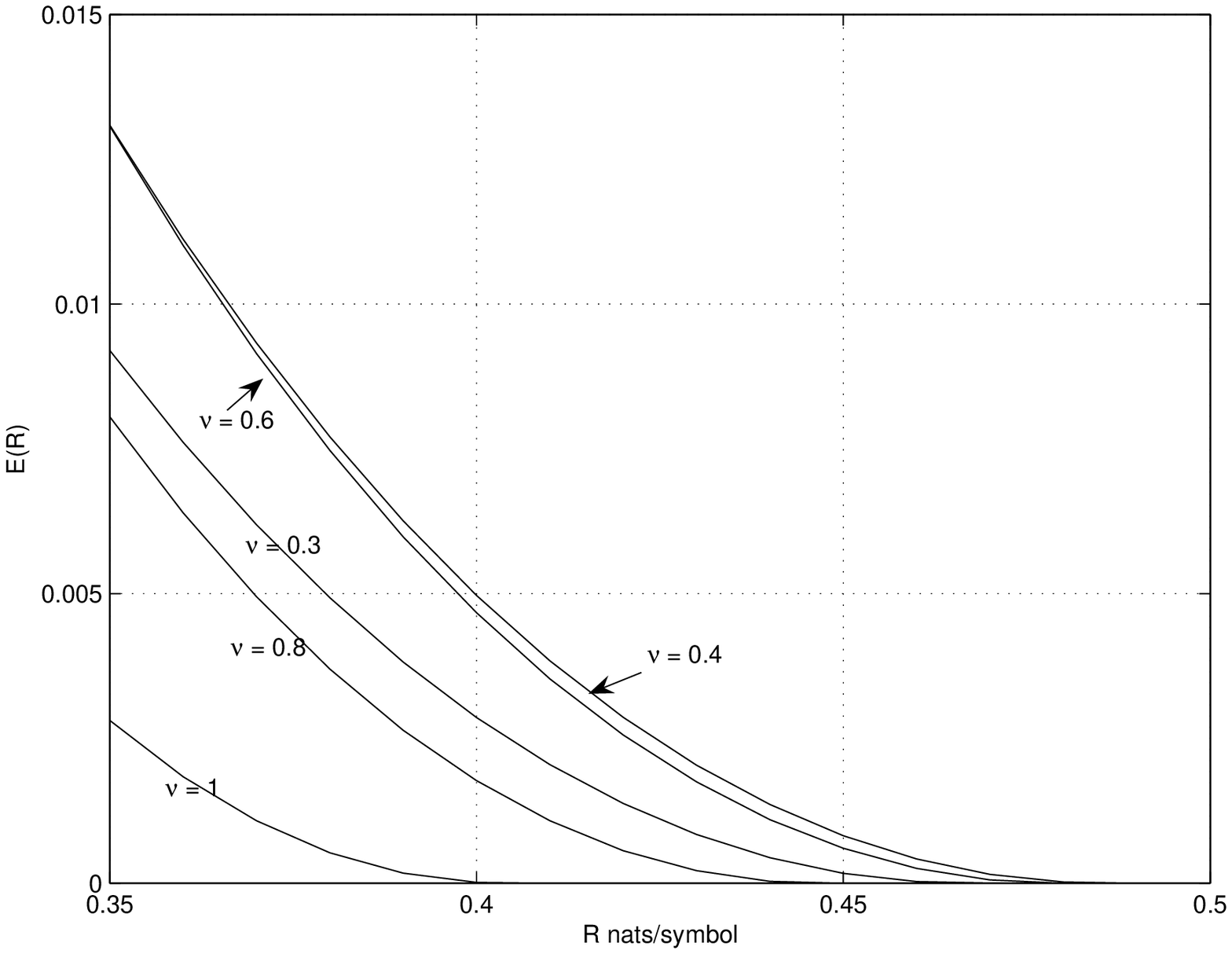}
\caption{Random coding error exponent of 16-OOPSK in the noncoherent
Rician fading channel with Rician factor $\K =
\frac{|d|^2}{\gamma^2} = 1$. The duty factor values are
$\nu=1,0.8,0.6, 0.4,0.3$. $\tsnr = 1$.}
\label{fig:eeoopsknM16highrates}
\end{center}
\end{figure}

\begin{figure}
\begin{center}
\includegraphics[width = 0.65\textwidth]{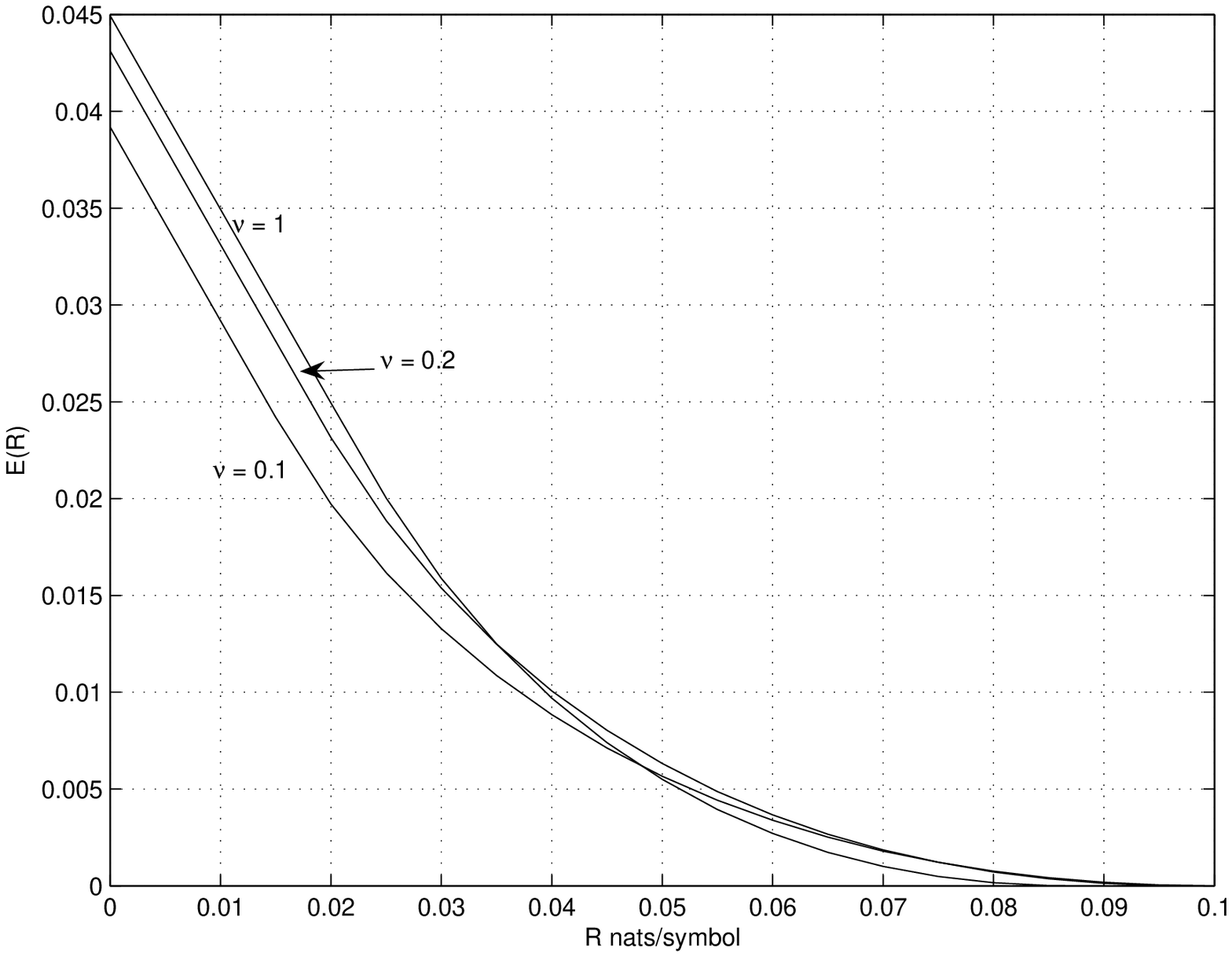}
\caption{Random coding error exponent of 16-OOPSK in the noncoherent
Rician fading channel with Rician factor $\K =
\frac{|d|^2}{\gamma^2} = 1$. The duty factor values are
$\nu=1,0.2,0.1$. $\tsnr = 0.1$.} \label{fig:eeoopsknM16snr01}
\end{center}
\end{figure}

\begin{figure}
\begin{center}
\includegraphics[width = 0.65\textwidth]{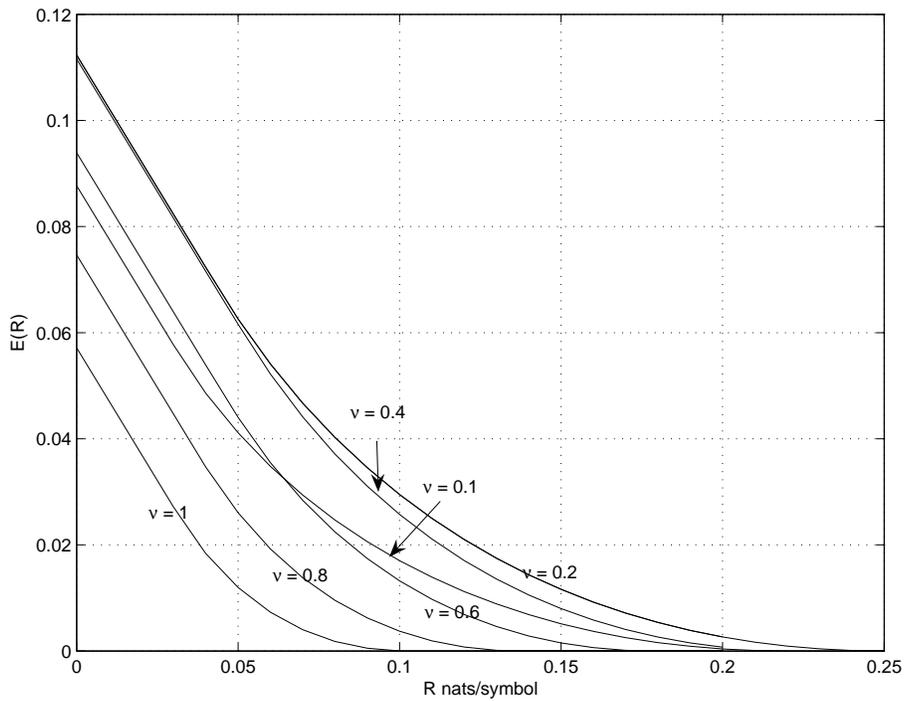}
\caption{Random coding error exponent of 2-OOFSK in the noncoherent
Rayleigh fading channel. The duty factor values are $\nu=1,0.8,0.6,
0.4,0.2,0.1$. $\tsnr = 1$.} \label{fig:eeoofsknM2}
\end{center}
\end{figure}

\begin{figure}
\begin{center}
\includegraphics[width = 0.65\textwidth]{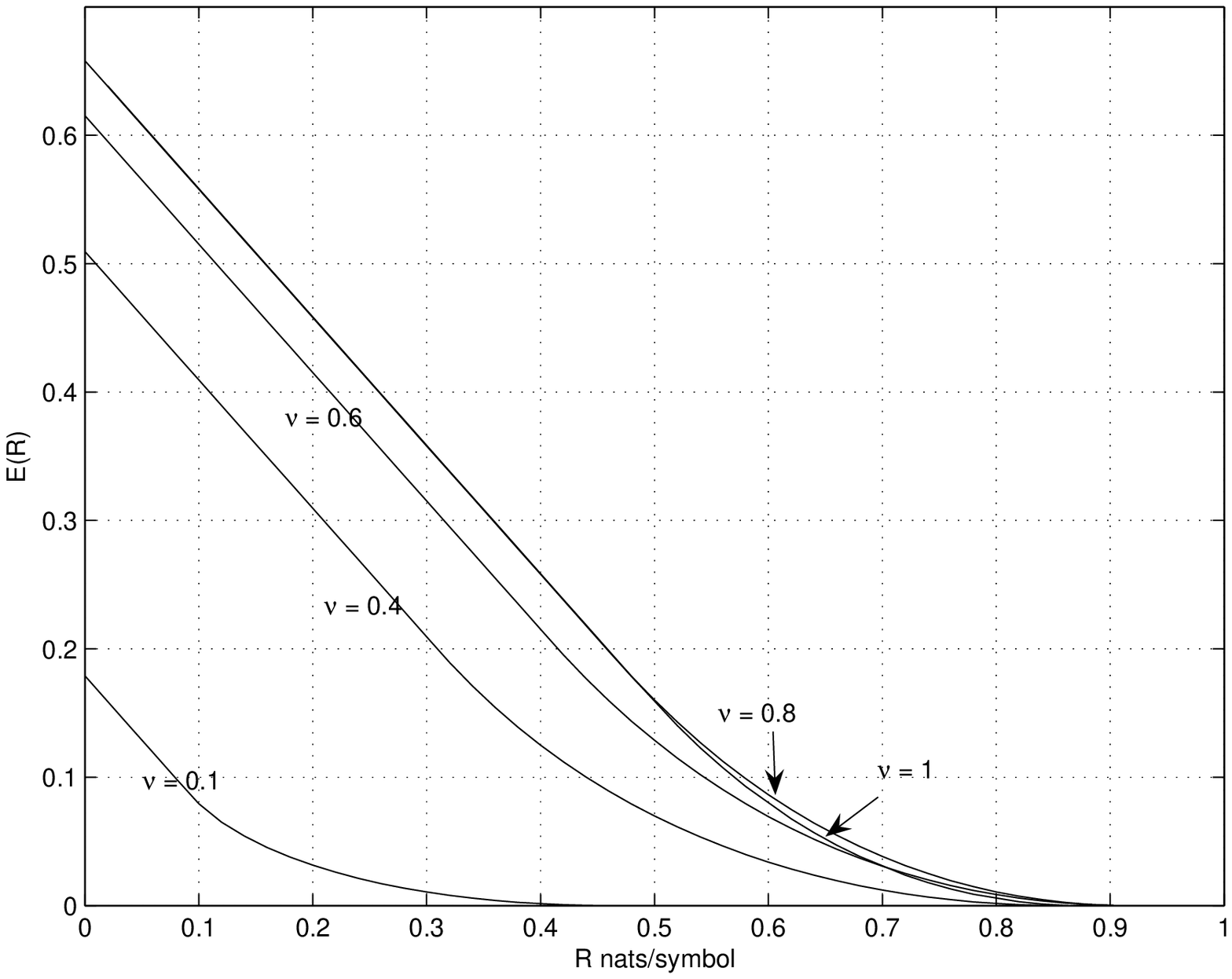}
\caption{Random coding error exponent of 16-OOPSK in the coherent
Rician fading channel with Rician factor $\K = 1$. The duty factor
values are $\nu=1,0.8,0.6, 0.4,0.1$. $\tsnr = 1$.}
\label{fig:eeoopskM16}
\end{center}
\end{figure}

We have numerically solved the optimization problem in
(\ref{eq:E(R)}) and we now present the results for the random coding
error exponents of OOPSK and OOFSK modulations in both coherent and
noncoherent fading channels. We first focus on noncoherent fading
channels. Fig. \ref{fig:eeoopsknM16} plots the error exponents as a
function of data rates $R$ of 16-OOPSK in the noncoherent Rician
fading channel with $\K = 1$ when $\tsnr = 1$. Fig.
\ref{fig:eeoopsknM16highrates} is a closer look at the exponents at
high rates. From these figures, we note that OOPSK with duty cycles
$\nu = 0.8,0.6,$ and 0.4 have higher error exponents with respect to
that of regular PSK (i.e., $\nu = 1$) over the entire range of rate
values. While signaling with $\nu = 0.6$ provides the highest
exponents for low rates, error exponents of $\nu = 0.4$ eventually
exceeds those of $\nu = 0.6$ at high rates as evidenced in Fig.
\ref{fig:eeoopsknM16highrates}. We further observe that the error
performance is relatively poor when $\nu = 0.1$. Fig.
\ref{fig:eeoopsknM16snr01} provides the error exponents in the same
channel when $\tsnr = 0.1$. We note that at this low value of the
$\tsnr$, signaling with low duty cycle (e.g., $\nu = 0.2$ or $0.1$)
results in improved performance at high rates in contrast to the
behavior at $\tsnr = 1$. Fig. \ref{fig:eeoofsknM2} plots the error
exponents of 2-OOFSK in the noncoherent Rayleigh fading channel when
$\tsnr = 1$. In this figure, we see that signaling with duty cycle
less than 1 improves the performance for all rates. Among the
parameters considered in the figure, $\nu = 0.2$ gives the highest
exponents at all rates. We have quiet different results when
coherent channels are considered. Fig. \ref{fig:eeoopskM16} gives
the error exponents of 16-OOPSK in the coherent Rician fading
channel with $\K = 1$. We immediately observe that in general
signaling with duty cycle less than 1 degrades the performance.
Improvements over the case of $\nu = 1$ is possible only at high
rates when $\nu = 0.8$. Hence, the gains in error performance that
we observe in uncoded systems are not realized in coded systems when
error exponents are considered.

\section{Conclusion} \label{sec:conclusion}

We have studied the error performance of peaky signaling over fading
channels. We have considered two modulation formats: OOPSK and
OOFSK. We have initially concentrated on uncoded systems. We have
found the optimal MAP decision rules and obtained analytical error
probability expressions for OOPSK and OOFSK transmissions over both
coherent and noncoherent fading channels. Through numerical results,
we have seen that error performance improves if the peakedness of
the signaling schemes are sufficiently increased. For fixed error
probabilities, substantial gains in terms of SNR per bit are
realized, making the peaky signaling schemes energy efficient. Since
decreasing the duty cycle diminishes the communication rates,
information-theoretic inspired OOPSK and OOFSK emerge as
energy-efficient modulation techniques well-suited for low data rate
applications. We have also analyzed the performance of coded
systems. We have numerically obtained the random coding error
exponents of both OOPSK and OOFSK. In noncoherent channels, we have
seen improvements in the performance if the duty cycles are less
than 1. On the other hand, we have observed that operating with low
duty cycles in coded systems in general degrades the performance in
coherent fading channels. This paper has mainly considered
single-user, single-antenna systems. Recently, we have considered in
\cite{QWang} the reception of OOFSK signals using multiple antennas
at the receiver. Results similar to those reported in this paper are
noted for the error performance. Future work includes the design of
peaky signaling schemes for multiple-input multiple-output (MIMO)
systems. Note that since signaling with a duty cycle decreases
collisions in a multiuser environment, the analysis of the impact of
peaky signaling on the design of medium access algorithms is also a
future research direction.

%\vspace{-2cm}

\end{spacing}

\newpage

\end{document}